\documentclass[preprint,12pt]{elsarticle}
\let\today\relax
\makeatletter
\def\ps@pprintTitle{%
    \let\@oddhead\@empty
    \let\@evenhead\@empty
    \def\@oddfoot{\footnotesize\itshape
         {} \hfill\today}%
    \let\@evenfoot\@oddfoot
    }
\makeatother

\usepackage{graphicx}
\usepackage{amssymb, amsmath}

\usepackage{lineno}
\usepackage{comment}
\usepackage{url}
\usepackage{caption}
\usepackage{subcaption}
\usepackage{float}
\usepackage{graphicx}
\usepackage{verbatim}
\usepackage{listings}
\usepackage{comment}

\begin{document}

\begin{frontmatter}


\title{Studying the UK Job Market During the COVID-19 Crisis with Online Job Ads}



\author{Rudy Arthur}

\address{Department of Computer Science, University of Exeter, North Park Road,Exeter,UK, EX4 4QF. \textit{E-mail address}: \texttt{R.Arthur@exeter.ac.uk}}

\begin{abstract}
The COVID-19 global pandemic and the lockdown policies enacted to mitigate it have had profound effects on the labour market. Understanding these effects requires us to obtain and analyse data in as close to real time as possible, especially as rules change rapidly and local lockdowns are enacted. In this work we study the UK labour market by analysing data from the online job board Reed.co.uk. Using topic modelling and geo-inference methods we are able to break down the data by sector and geography. We also study how the salary, contract type and mode of work have changed since the COVID-19 crisis hit the UK in March. Overall, vacancies were down by 60 to 70\% in the first weeks of lockdown. By the end of the year numbers had recovered somewhat, but the total job ad deficit is measured to be over 40\%. Broken down by sector, vacancies for hospitality and graduate jobs are greatly reduced, while there were more care work and nursing vacancies during lockdown. Differences by geography are less significant than between sectors, though there is some indication that local lockdowns stall recovery and less badly hit areas may have experienced a smaller reduction in vacancies. There are also small but significant changes in the salary distribution and number of full time and permanent jobs. In addition to these results, this work presents an open methodology that enables a rapid and detailed survey of the job market in these unsettled conditions and we describe a web application \url{jobtrender.com} that allows others to query this data set.
\end{abstract}


\end{frontmatter}

\section{Introduction}
\label{S:1}

The COVID-19 pandemic claimed over 90,000 lives in the UK as of the January 2021 \cite{deaths}. The most drastic measure to limit the spread of COVID-19 has been the imposition of so-called lockdown measures. `Lockdown' refers to national or regional orders calling for the closure of businesses and restriction of assembly and travel. Lockdown policies have not been uniform across nations. Some areas, for example Sweden and Japan, implemented essentially voluntary measures \cite{Sweden, Japan} while others, for example China and Germany, imposed and enforced quite severe restrictions on assembly and business opening \cite{Germany, China}.

The UK's lockdown policy was between these two extremes. Beginning somewhat later than many other European nations on March 21st the UK government introduced the The Health Protection (Coronavirus, Business Closure) (England) Regulations 2020 \cite{uklockdown1} which was superseded by the The Health Protection (Coronavirus, Restrictions) (England) Regulations 2020 on March 26th \cite{uklockdown2}. This piece of legislation, which we will hereafter refer to as `lockdown', included restricted freedom of movement, bans on gatherings and enforced business closures. The lifting of some of these rules began on May 13th, though in some heavily affected areas stricter measures were retained or re-imposed \cite{blackburn, leicester} and the rules have been subsequently modified e.g. `the rule of six' \cite{uklockdown3}, the `tier system' and other national lockdowns \cite{tiers}.

The economic impacts of the COVID-19 crisis have been severe. The UK saw a 125.9\% increase in unemployment claims between March and May and vacancies dropping by 58\% over the same period \cite{ferreira2020iza}. Headline unemployment numbers did not immediately rise \cite{unemployment} but by the end of the year the employment rate was on a markedly downward trend, standing at 4.9\% as of December 2020 \cite{unemploymentDec}. To provide support during this period the UK government introduced unprecedented measures: namely the Coronavirus Job Retention (furlough) Scheme \cite{furlough} to attempt to keep unemployment rates in check by providing grants to pay up to 80\% of salaries.

The academic study of lockdown has necessarily been re-active and observational. The rapid onset of the crisis has meant that researchers have had to source and analyse real time labour market information. For example Bick et. al \cite{bick2020real} use an online labour market survey; Chetty et. al. \cite{chetty2020real} use anonymised data from several large companies; Hensvik et. al. \cite{hensvik2020job} use vacancy postings while Forsythe et. al \cite{kahn2020labor} use data from the job market analysis company Burning Glass (\url{burning-glass.com}). 

This methodology yields useful insights into the effect of the COVID-19 crisis on the labour market. For example \cite{kahn2020labor} show a 44\% drop in vacancy postings between February and April in the US, observed across occupational categories (essential or non-essential work) as well as states which may have had different lockdown policies.
Bick et. al. \cite{bick2020real} report similar drops in the US across sector and demographics during the same period, with a slow recovery in the months after. Analysis of vacancy postings in Sweden, which has had probably the least restrictive lockdown in Europe, shows a drop in job adverts by around 40\% \cite{hensvik2020job}. Hensvik et. al. also report that job seekers are searching less intensively and redirecting their searches towards less severely hit occupations. 

The effect of the crisis on different demographic sectors of the economy has exacerbated previously existing inequalities \cite{montenovo2020determinants}. In Europe Adams-Prassl et. al. \cite{adams2020inequality} found that workers in Germany were insulated from job losses by longstanding institutional frameworks, compared to workers in the UK who are in a much more precarious position. The same work also finds that job losses and reductions in earnings disproportionately affect women, workers without a university degree and younger people.  

 There is still serious academic debate about the efficacy of lockdown as a disease control measure \cite{ioannidis2020coronavirus,chaudhry2020country}, how it should be implemented \cite{islam2020physical} and the negative and unintended consequences of these policies \cite{williams2020improved} on health outcomes. This work is an observational study of the effect of COVID-19, lockdown and other disease control measures on the job market in the UK. We will use topic modelling and location inference methods on job vacancies to cross section the data aiming to provide insight into the magnitude of the effect of COVID-19 and associated policies on job vacancies across sectors and geographies in the UK.
 
The technical approach we take is similar to that used by Burning Glass or \cite{thurgood2018using}. In \cite{thurgood2018using} Turrell et. al. used an unsupervised machine learning method on a corpus of online job adverts to discover how the labour market is segmented. In this work we use a similar data set but apply a new supervised method which gives more robust results. We also add a location inference step to enable geographic analysis. Finally we will briefly describe a web application where this data can be queried by interested parties. We hope that the novel topic modelling and geographic inference methods coupled with an interactive web interface make this a useful tool for academic researchers and government agencies looking to understand the `live' job market.

\section{Data and Methods}
\label{S:2}

We collect jobs from the online job board Reed.co.uk. Reed is a large recruitment agency and owner of the UK's first recruitment website, which recieves around 7 million visits per month (\url{reed.co.uk/about}). Apart from its popularity, what makes it suitable for this analysis is an API (\url{reed.co.uk/developers}) which allows for download of job adverts. 

Each job advert on Reed has a unique sequential index number. We start at index 37000000, with our first full day of collection January 11th 2019. We use two full years of data, up to January 11th 2021, during which we collected around 4.7 million job ads. This should be sufficient to establish a baseline of vacancy information before COVID-19 became a national issue, observe the entire first lockdown period and national reaction as well as a long enough time-frame to study any potential recovery. With only two years of data we cannot observe long term trends, for example due to Brexit or other social, political and economic forces, for that we rely on official statistics e.g. \cite{unemploymentDec}.

\begin{figure}
\centering
\includegraphics*[width=\textwidth]{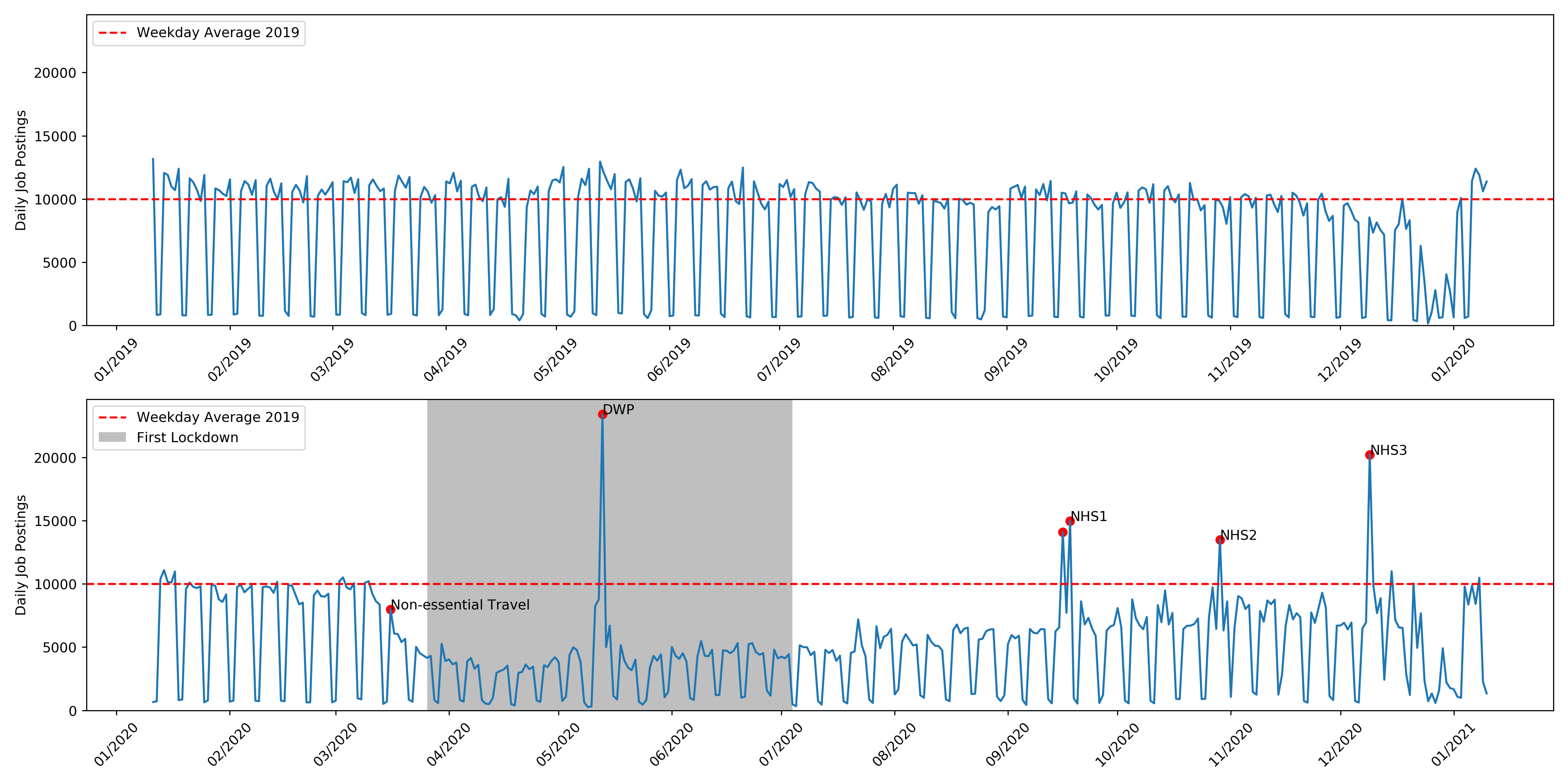}
\caption{Daily Job postings on the job board over the study period. The periodic dips correspond to weekends and public holidays. Some key dates are indicated and discussed in the text. }\label{fig:total}
\end{figure}

Figure \ref{fig:total} shows the number of job adverts collected per day. Job adverts which have been deleted or removed are returned by the API as JSON (JavaScript Object Notation) objects with all null values and are not included in Figure \ref{fig:total}. There is no indication that these removed job ads have significant impact and they represent only around 3\% of the total number of records, spread fairly uniformly across the period. 

Figure \ref{fig:total} shows some trends quite clearly. We have indicated some key events on the plot, as well as the period we have defined as `lockdown'. The 2019 data shows a remarkably steady number of daily job ad postings through the year, however there is a distinct downward trend beginning in late November/early December. We can see that the first recorded case of coronavirus in the UK on the 31st of January 2020 is not associated with any change in trend. At this point coronavirus was not considered to be a major issue in the UK. A distinct drop in the number of job ads posted starts on March 16th when UK Prime Minister Boris Johnson issued advice against ``non-essential'' travel and contact. Voluntary measures were in place before this as well as press rumours of the soon to be announced lockdown and the data indicates the start of the drop in job vacancies preceded the announcement. By the time lockdown began the number of ads was reduced to under half of the early year baseline. 

A major spike in job ads, labelled DWP, occurs on the same day as  the second amendment to the Health Protection Act \cite{uklockdown2}, 13/5/2020, which allowed for the re-opening of certain businesses and services e.g. garden centres, tennis courts and recycling centres. However we cannot conclude that this spike was caused by the announcement, as the vast majority of these jobs were posted by the UK's Department of Work and Pensions. This body, among other things, manages the UK's welfare system and maintains its own job advertisement board, \url{findajob.dwp.gov.uk}. On this day roughly 20000 job ads of all kinds were cross-posted to Reed from this site. Other spikes are labelled NHS1, NHS2 and NHS3 also correspond to a sudden surge in job advertisement from government agencies. These spikes represent between 8000 and 14000, mostly nursing, jobs posted by the NHS Business Services Authority on the corresponding days. This agency also has its own job board, \url{jobs.nhs.uk}, and these are likely cross postings designed to give it a wider reach.

\begin{figure}
\centering
\includegraphics*[width=\textwidth]{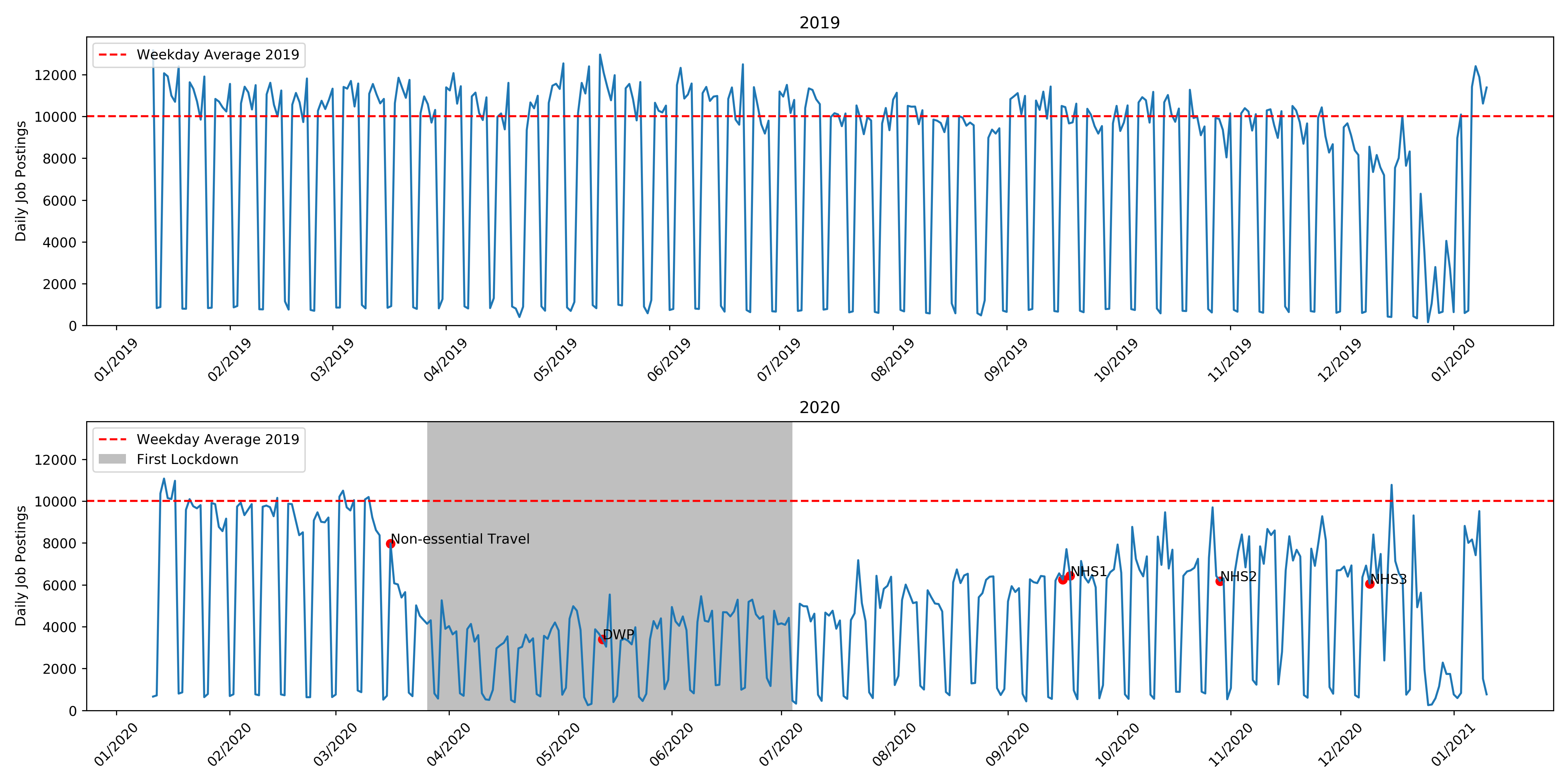}
\caption{Daily job postings with the cross posts from the two other major job boards removed.}\label{fig:cleantotal}
\end{figure}

In light of this, we remove jobs posted by these two agencies from our data set, that is remove all ads posted by employers DWP Teaching, Department of Work \& Pensions and NHS Business Services Authority. The resulting time series is shown in Figure 2. With the spikes removed we can see the number of jobs posted per day rapidly fell in March but has since slowly grown from its nadir in mid-April. As of January 2021 the number of posts per day seems to have recovered to pre-crisis levels. The total number of job ads posted in the 2019 period was 2691648 compared to 1700687 in the 2020 period, a deficit of 990961 or 37\% fewer job adverts. If we take March 16th 2020 as the date at which the job market began to react to the crisis comparing March 16th to Jan 10th in both periods gives 2166551 and 1245180 job ads in 2019 compared to 2020, a deficit of 42.5\%.

\subsection{Topic Modelling}

Though the main site is searchable by topic and sector, the JSON payload returned by Reed's API does not include a theme or topic marker and so this must be inferred. Algorithms such as LDA \cite{blei2003latent} and Doc2Vec \cite{le2014distributed} transform text documents to low dimensional vector representations which enhance automatic topic detection algorithms. LDA has previously been used with success on a very similar data set \cite{thurgood2018using}. We attempted to model our data using both of these methods. 

While some categories of job are readily detectable with unsupervised methods e.g. software developer, the topics or clusters detected are often not stable when varying the algorithm parameters. Measures like coherence \cite{roder2015exploring} fail to provide obvious evidence in favour of any one parameter set and in particular an optimal number of topics. While we attempted unsupervised clustering, namely LDA and Word2Vec, we do not use unsupervised methods in this work. A second reason to avoid unsupervised methods is that we want to study predetermined categories of job e.g. nursing, teaching or graduate jobs, to observe the reaction of these sectors to the COVID crisis. Thus we know the labels we want and should be able to obtain better results by providing more information to the classification algorithm. We therefore take a different and somewhat simpler approach which nevertheless suffices to identify topics.

\begin{table}
    \centering
    \begin{tabular}{cccc}
account & accountant & administrator & assistant \\ business & buyer & care & charity \\ 
cleaner & construction & customer & data \\ 
delivery & electrician & finance & garage \\ 
graduate & hgv & hotel & hr \\ 
itsupportengineer & kitchen & machine & marketing \\ nurse & nursery & physio & plumber \\ 
prison & production & productionmanager & project \\ property & receptionist & recruitment & retail \\ 
sale & security & server & software \\ 
solicitor & storemanager & support & surveyor \\ 
teacher & vehicle & warehouse & welsh \\ 
other
    \end{tabular}
    \caption{Job category labels. These labels are summaries of the most common job title in each category. The category `other' contains jobs which could not be classified into one of the other categories.}
    \label{tab:job_cat}
\end{table}
For each ad we combine the job title with the job description to constitute a `document'. The document text is cleaned to remove HTML artefacts, lower-cased, lemmatised using the wordnet lemmatiser \cite{wordnet} and tokenised. Using data from 2019 (Reed job ids 39500000 to 39600000) we collect a number of `seed' documents. These seeds are collections of job ads which are representative of given sectors $S_c = \{ d_{c1}, d_{c2}, \ldots, d_{cn} \}$. The label $c$ is a job category e.g.` teacher' and the documents $d_{ci}$ are job adverts definitively in that sector. These were determined by searching for common job titles in each category e.g. in `retail' we look for jobs with the terms `order picker', `shop supervisor', `retail assistant' or `picker packer' in the job title. The full list and the jobs used to build the seeds is available in the linked git repository \ref{sec:SI}. These categories were obtained by manual inspection of the data as well as comparison with the browsable job areas on the main Reed website.

To get examples of the `other' category we used Gensim's implementation of TF-IDF \cite{rehurek_lrec} to transform each document into a vector. We compute the mean cosine similarity of each job with all the vectors in each seed set. If the job has similarity score less than 0.04 (roughly the 5th percentile) with any of the categories we use it as an example of the `other' category. In this way we identify 2205 jobs as not similar to any of the given classes and label them as `other'. The number 0.04 was found to work well in practice, however the method is robust to variations of this threshold.

With the given seeds we tried a number of classification methods, finding that a simple decision tree \cite{scikit-learn} preformed best, achieving a Subset Accuracy score of 0.922, a Balanced Accuracy of 0.884, and a Cohen's Kappa of 0.919 on an 80/20 train/test split of the seed data. When we use the classifier on 100000 different job ads we find that 85\% are classified into one of the named categories with 15\% put into the `other' class. We will show word clouds built from documents in a number of classes in section \ref{sec:sector}, these seem very reasonable and a manual inspection of the confusion matrix finds that mistakes are almost invariably confusion of two similar categories e.g. `accountant' and `finance'. Table \ref{tab:misclas} shows the most common mis-classifications.

\begin{table}
    \centering
    \begin{tabular}{|c|c|}
    \hline
    True/False & Count \\
garage/customer &  5  \\
sale/customer &  5  \\
finance/hr &  5  \\
account/accountant &  5  \\
business/sale &  6  \\
account/finance &  8  \\
administrator/finance &  9  \\
finance/accountant &  26  \\ \hline
    \end{tabular}
    \caption{Out of 8844 examples in the test set 196 are misclassified. The most common confusion made by the decision tree seems to be between office finance roles e.g. `finance' includes jobs like `finance assistant' and `purchase ledger clerk' which have similar requirements to `accountant' jobs. Many of these jobs are classified similarly by Reed themselves, so the small number of classification errors we make are on potentially ambiguous ads, showing the strength of the method.}
    \label{tab:misclas}
\end{table}

\subsection{Location Inference}

Location inference is done using the `location' field in the advert's JSON. Some adverts are only localised at the county level, e.g. Devon. These are identified by checking against a list of UK administrative counties. The rest of the adverts are checked by querying the location field against the Geonames \cite{GeoNames} and Nominatim databases \cite{Nominatim}, in that order. Geonames is effective at returning co-ordinates for larger towns and cities \cite{arthur2018social}. Nominatim is a web based geo-coding service which uses OpenStreetMap data to find locations by name and address. Since the same locations re-occur multiple times, every Nominatim lookup is saved in a database which is queried before calling the web service. This process returns GPS coordinates or bounding boxes for 97\% of all non-null job adverts. 
\begin{figure}[H]
    \centering
        \includegraphics[height=0.88\textheight]{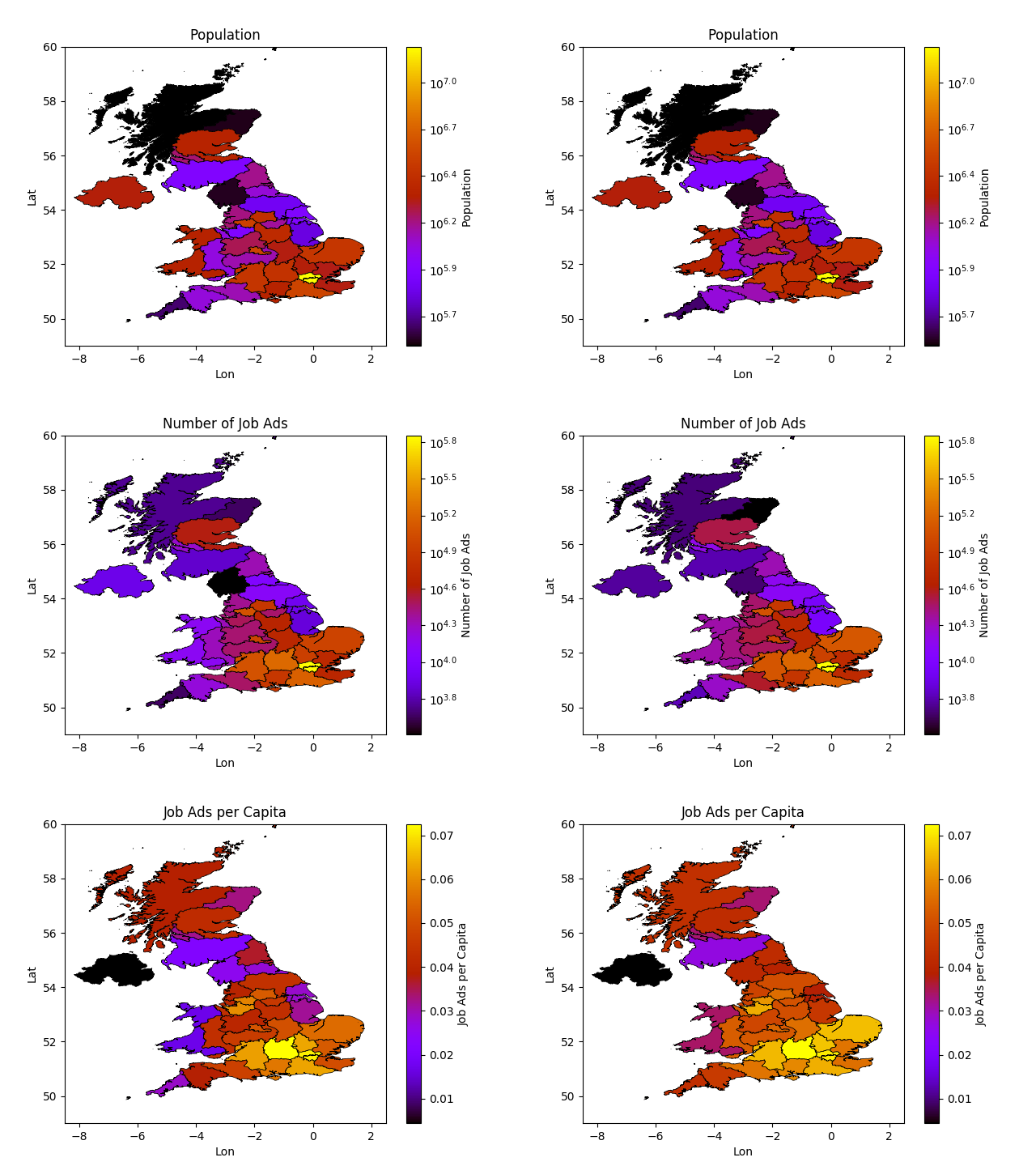}
     \caption{ The administrative regions shown are the NUTS2 regions, except the London NUTS2 regions have been combined into `Greater London'. This is because a very large number of job ads give the location `London', rather than a specific borough. The top row shows UK population, the middle shows the number of job ads posted and the bottom plot shows the number of job ads divided by the population of each region. The left column is 2019 and the right is 2020.}
     \label{fig:maps}
\end{figure}
Figure \ref{fig:maps} shows the spatial distribution of job ads, localised to NUTS2 regions \cite{nuts}. The  geographic distribution of job adverts roughly corresponds to the UK population distribution \cite{ukpopulation}. The number of ads normalised by population (the bottom maps in figure \ref{fig:maps}), shows a higher density in the south east. There are also some changes from 2019 to 2020, despite over all fewer jobs the per capita numbers are slightly higher in north east of England and rural areas of Cornwall and west Wales. The north/south divide is likely reflective of economic disparities within the UK \cite{Giovannini2019DividedAC}. However there are also potentially differences in the popularity of the job board Reed.co.uk in different regions, or for different industries, e.g. it hosts very few farming jobs. Even without knowing how use of Reed is spread across the UK we can still look at relative trends. Some changes may reflect a drift in the user base of the site, but large and sudden changes are likely to be caused by significant exogenous events.
\section{Jobs by Sector}
\label{sec:sector}

This section presents a collection of time series for different job classes identified by our topic modelling approach over the study period. These classes were chosen to highlight job categories which were likely to be impacted differently by the COVID-19 crisis. We expected delivery, nursing and care work demand to increase, while we expected hospitality and graduate jobs to decrease, the latter expectation informed by \cite{adams2020inequality}. Software was chosen as a job which could in principle be done remotely so it would be interesting to see how this affected vacancy numbers. Finally we expected the number of teachers required to be largely unaffected by the crisis. 

These time series are accompanied by word clouds constructed from the titles of the matched job adverts to give a sense of the job category. The captions of the figures will give some commentary on the observed trends. The job categories have been chosen to be representative of some key sectors and to show interesting trends, they do not represent every job sector that can be found in the data. When we construct word clouds we only use jobs from the period March 16th to Jan 10th in both years. This is to attempt to detect any change in common job titles caused by COVID. \clearpage

\begin{figure}[H]
     \centering
     \begin{subfigure}[H]{\textwidth}
         \centering
         \includegraphics[width=\textwidth]{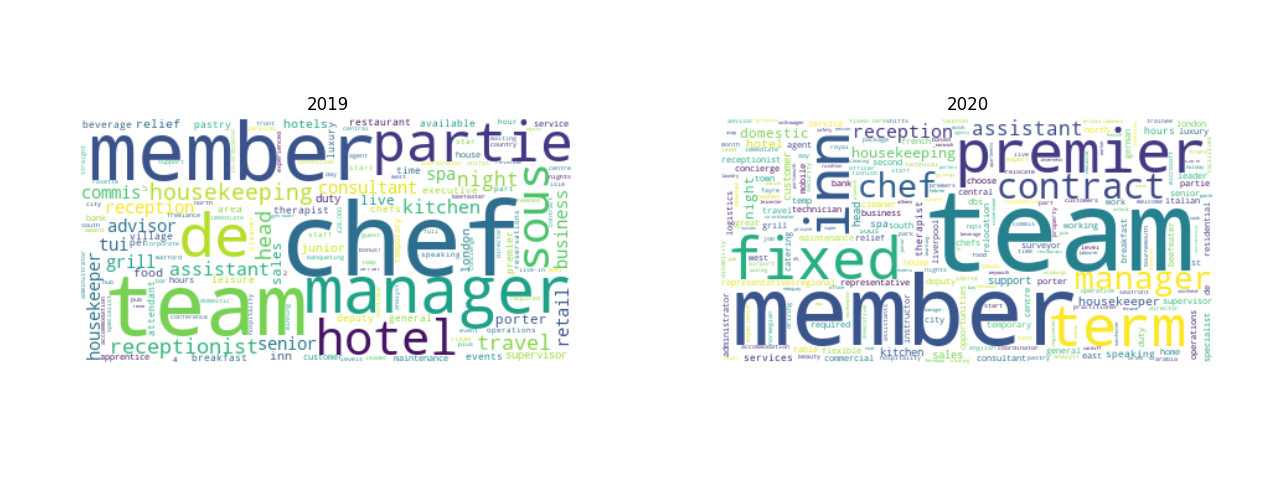}
     \end{subfigure} 
     \begin{subfigure}[H]{\textwidth}
         \centering
         \includegraphics[width=\textwidth]{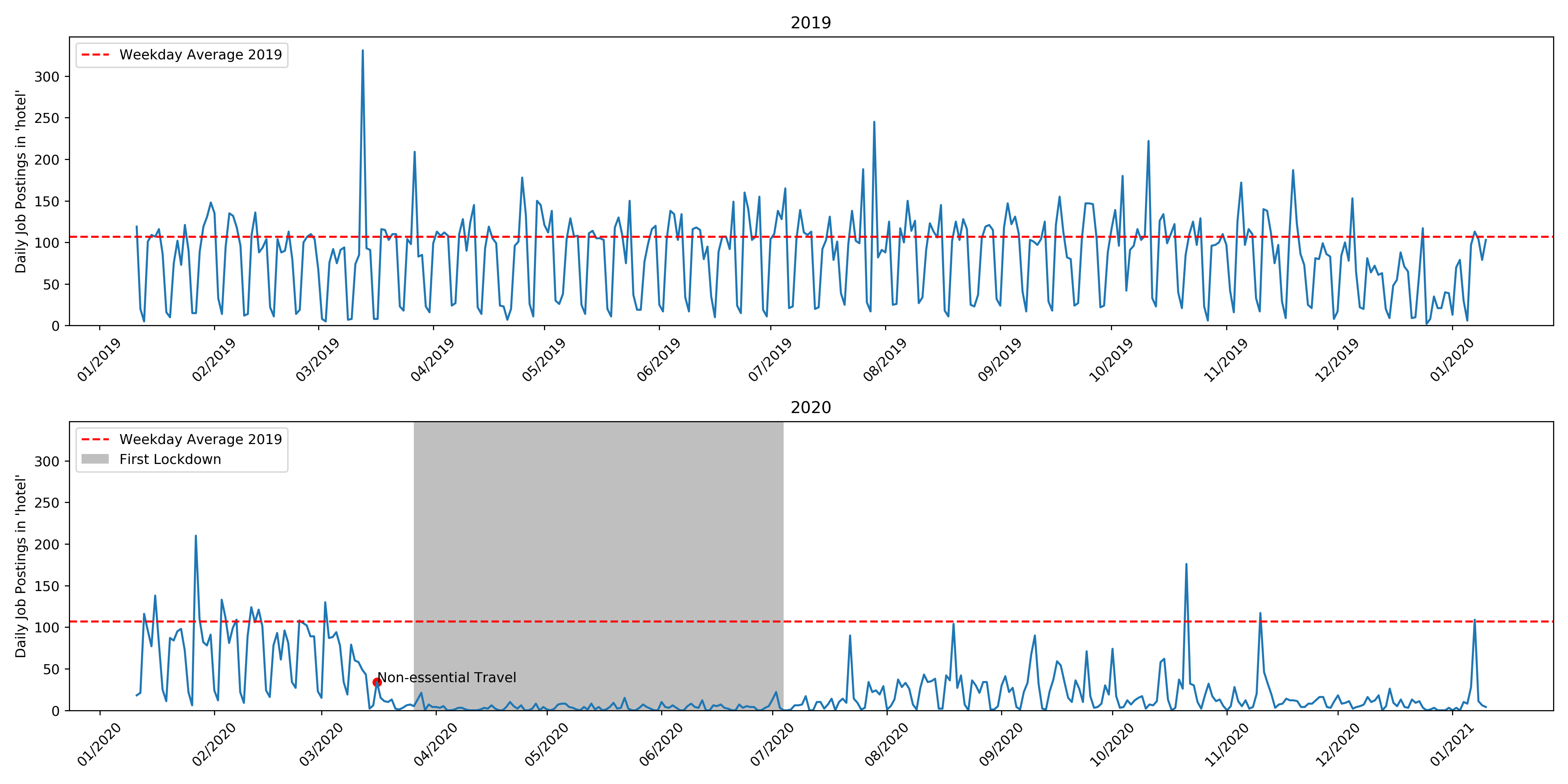}
     \end{subfigure}
        \caption{Jobs matching `hotel'. This term collects a number of service industry jobs in the hotel and hospitality industry. We see a complete collapse in the number of vacancies in this sector in the wake of the COVID-19 crisis with the drop beginning the week prior to the first major announcement of lockdown restrictions. The number of vacancies remained low throughout lockdown and in the weeks after. A recovery began in mid-August, summer holiday season, but vacancy postings remain at less than half of pre-crisis levels. Interestingly in the word cloud for 2020 we see `fixed term' indicating those jobs that are being advertised are not permanent. }
        \label{fig:hotel}
\end{figure}

\begin{figure}[H]
     \centering
     \begin{subfigure}[H]{\textwidth}
         \centering
         \includegraphics[width=\textwidth]{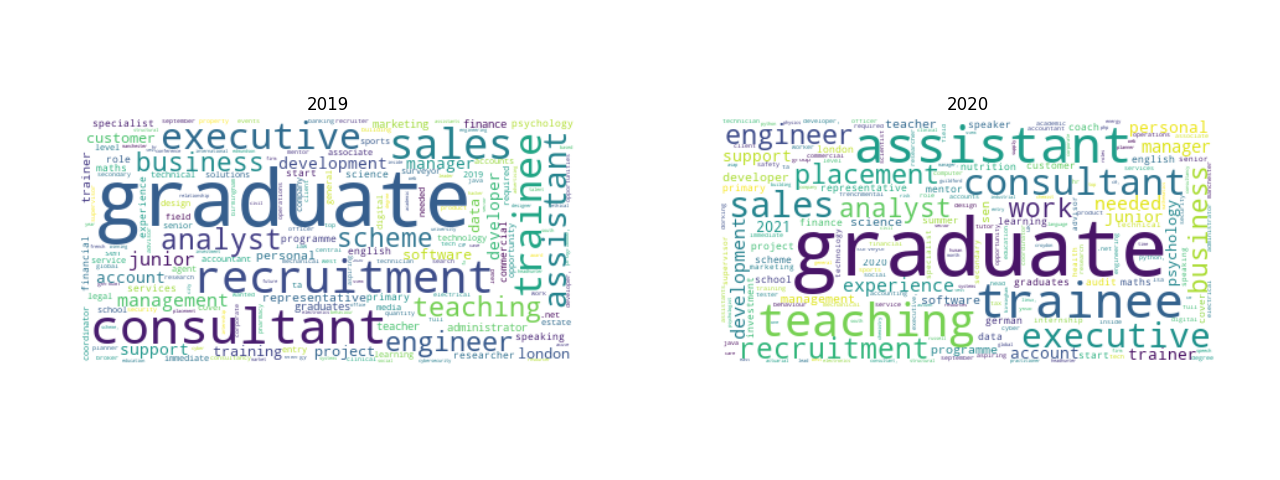}
     \end{subfigure} 
     \begin{subfigure}[H]{\textwidth}
         \centering
         \includegraphics[width=\textwidth]{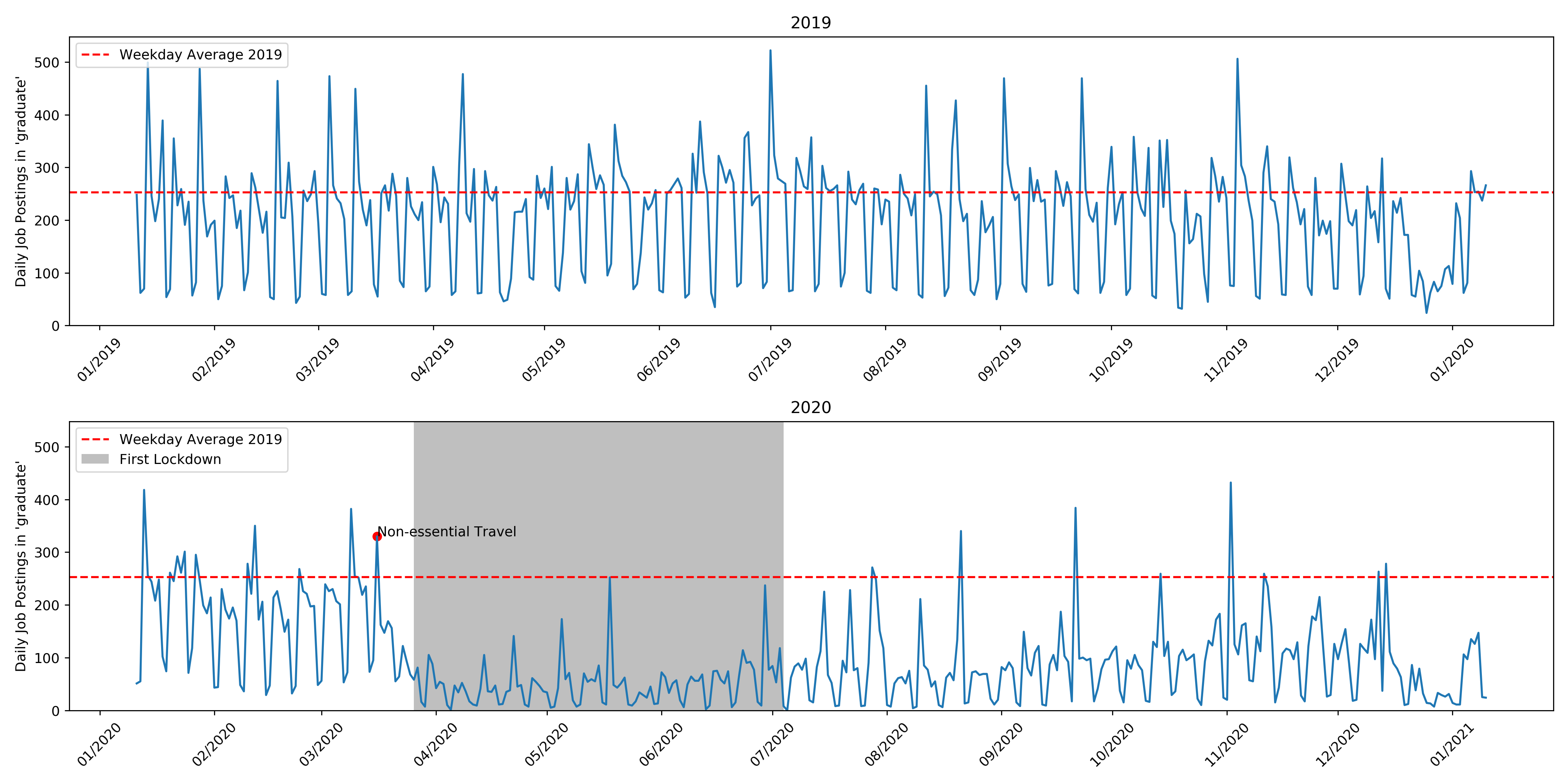}
     \end{subfigure}
        \caption{Jobs matching `graduate'. This cluster contains vacancies for graduates entering the workforce via job training or recruitment schemes. In line with \cite{adams2020inequality} we find a drastic drop in the number of adverts for these schemes, with a very slow recovery in the weeks after lockdown. }
        \label{fig:graduate}
\end{figure}

\begin{figure}[H]
     \centering
     \begin{subfigure}[H]{\textwidth}
         \centering
         \includegraphics[width=\textwidth]{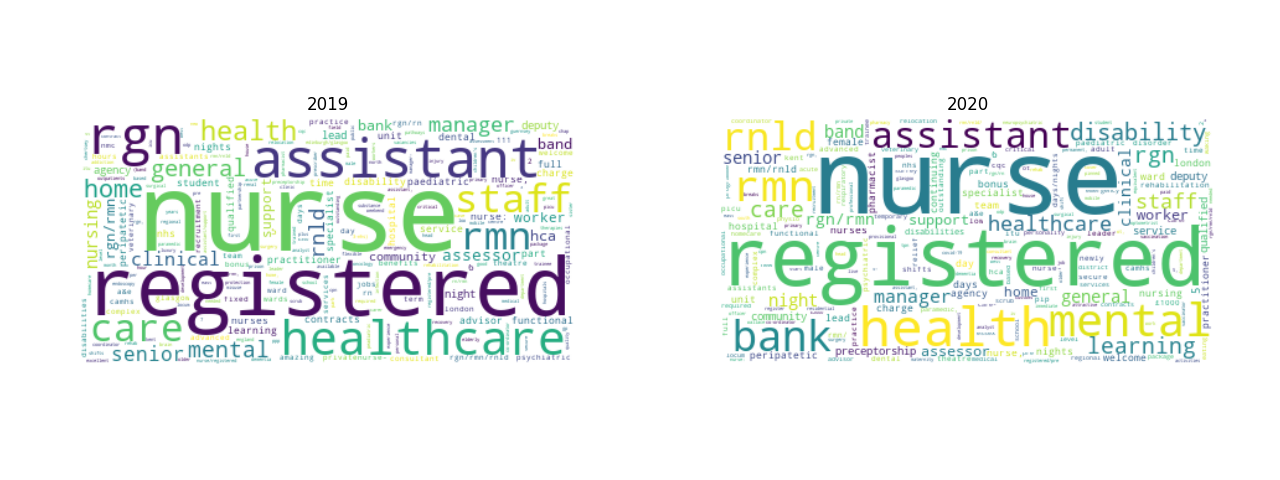}
     \end{subfigure} 
     \begin{subfigure}[H]{\textwidth}
         \centering
         \includegraphics[width=\textwidth]{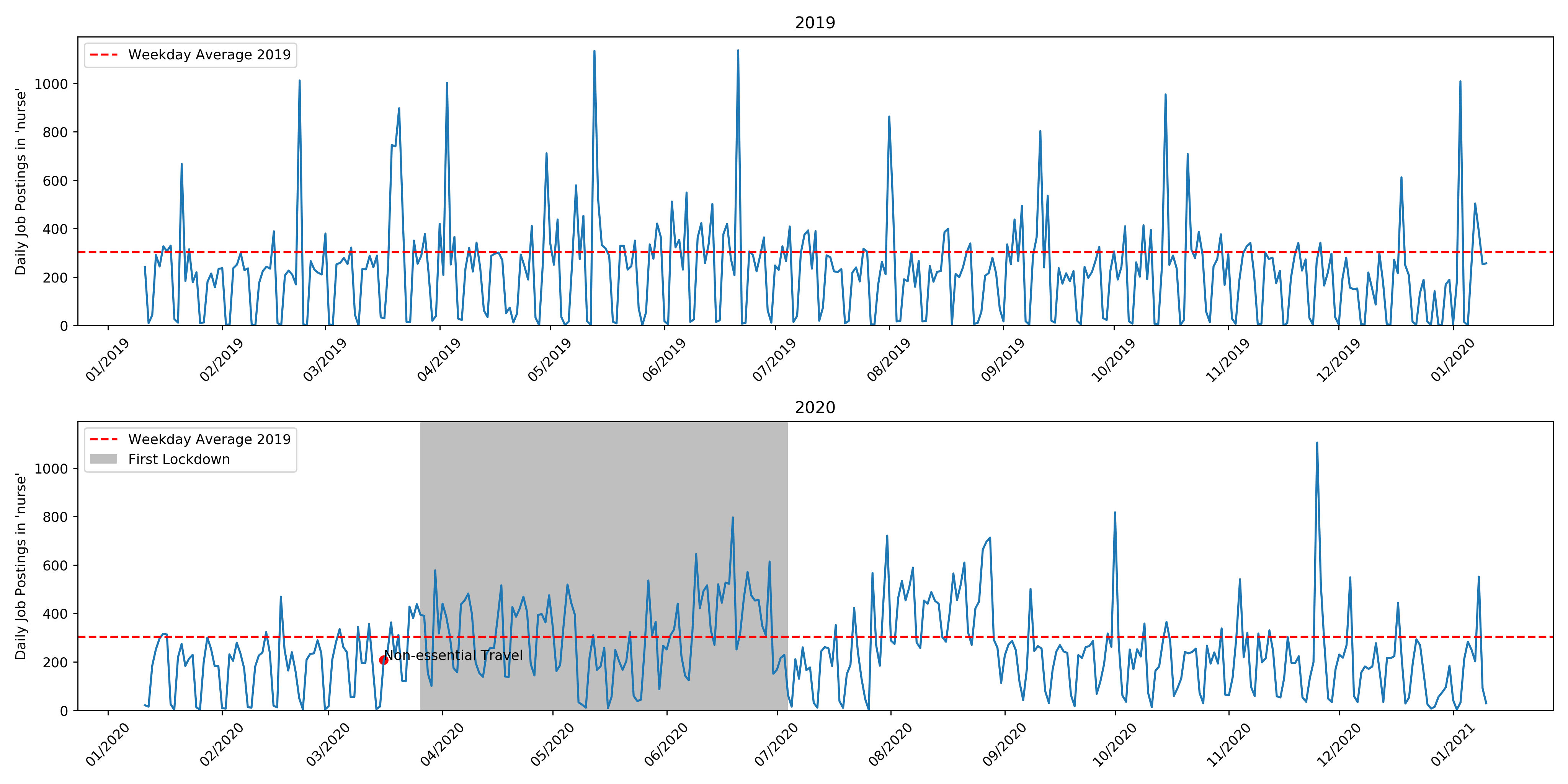}
     \end{subfigure}
        \caption{Jobs matching `nurse'. As might have been expected during a public health emergency, the number of ads for nurses increased somewhat during lockdown and in the weeks after. The word clouds for 2020 shows a high occurrence of the terms `mental health'. The mental health implications of COVID-19 have been much discussed \cite{pfefferbaum2020mental} and hiring decisions are reflecting this. }
        \label{fig:nurse}
\end{figure}

\begin{figure}[H]
     \centering
     \begin{subfigure}[H]{\textwidth}
         \centering
         \includegraphics[width=\textwidth]{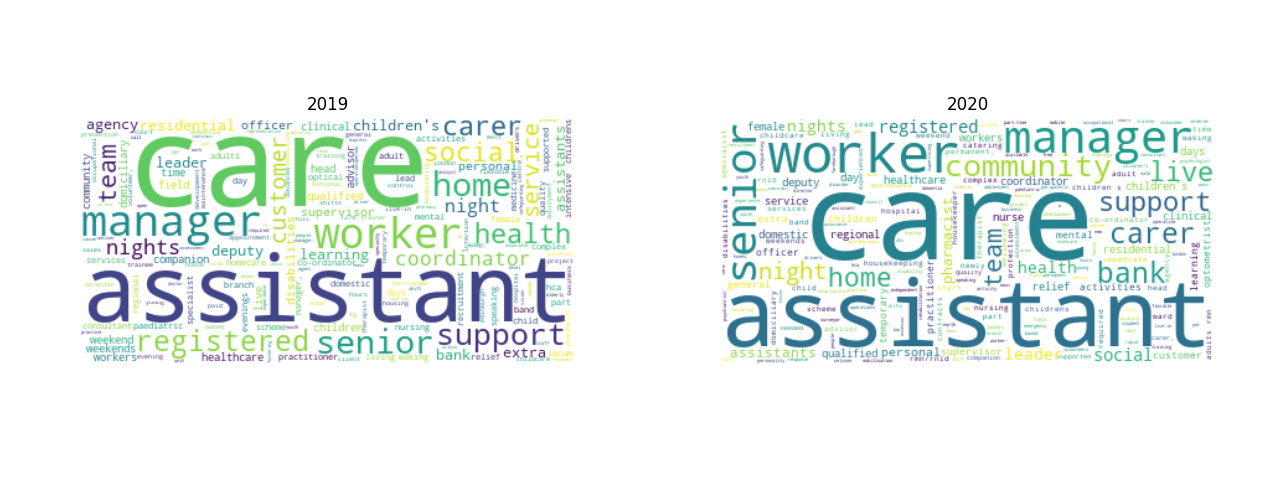}
     \end{subfigure} 
     \begin{subfigure}[H]{\textwidth}
         \centering
         \includegraphics[width=\textwidth]{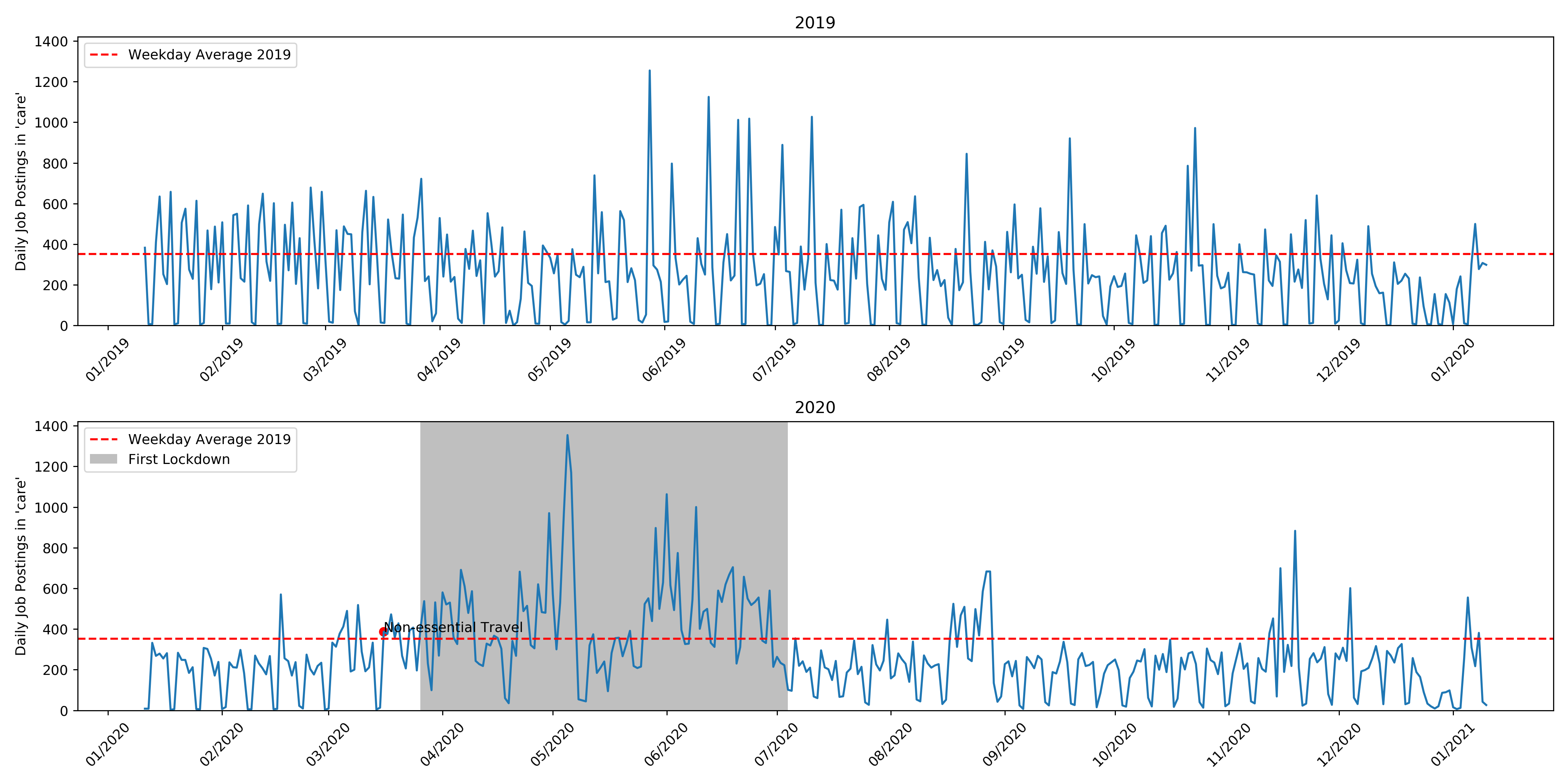}
     \end{subfigure}
        \caption{Jobs matching `care'. This time series also shows an increased demand for care workers during the crisis, especially during the lockdown period. The closure of many services through this period likely increases the need that older or disabled people have for special assistance.}
        \label{fig:careworker}
\end{figure}

\begin{figure}[H]
     \centering
     \begin{subfigure}[H]{\textwidth}
         \centering
         \includegraphics[width=\textwidth]{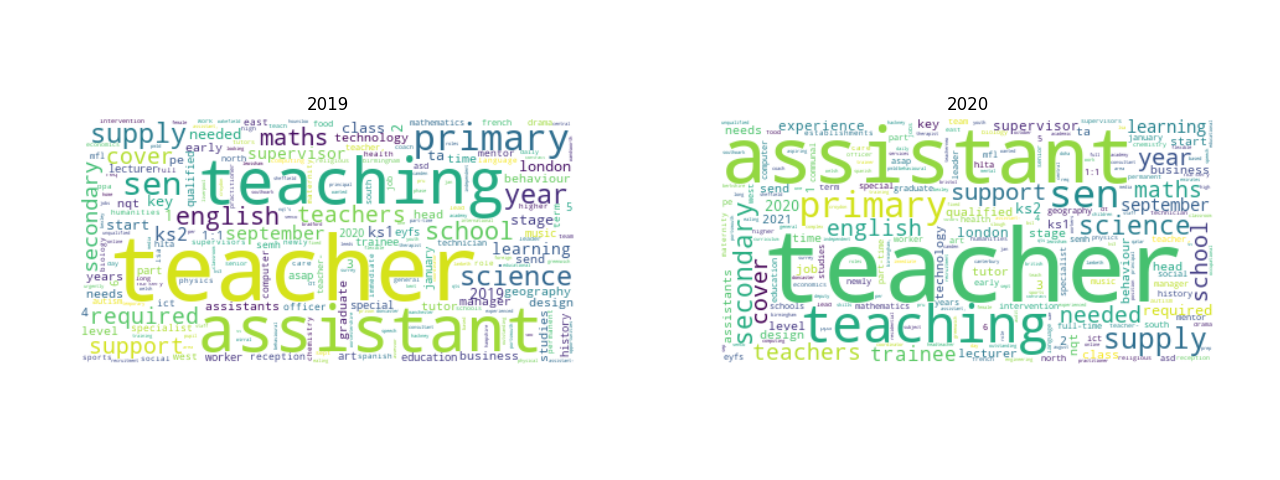}
     \end{subfigure} 
     \begin{subfigure}[H]{\textwidth}
         \centering
         \includegraphics[width=\textwidth]{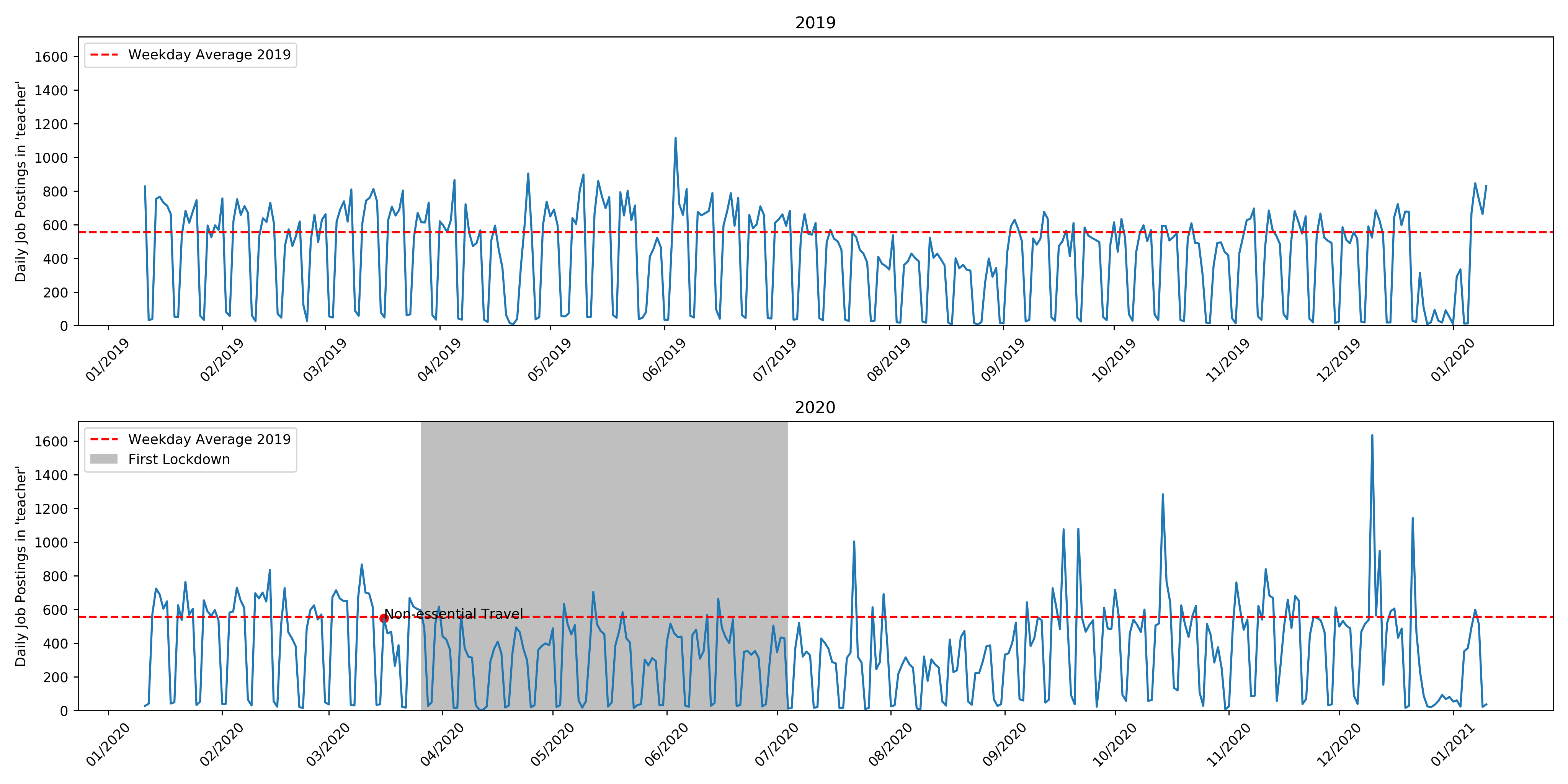}
     \end{subfigure}
        \caption{Jobs matching `teacher'. This time series shows a slight decrease in the demand for teachers during lockdown, however some of the backlog seems to have been filled once the first easing of restrictions was announced. After lockdown the number of vacancies in this sector is still down slightly, but is reduced far less than other sectors and the average. }
        \label{fig:teacher}
\end{figure}

\begin{figure}[H]
     \centering
     \begin{subfigure}[H]{\textwidth}
         \centering
         \includegraphics[width=\textwidth]{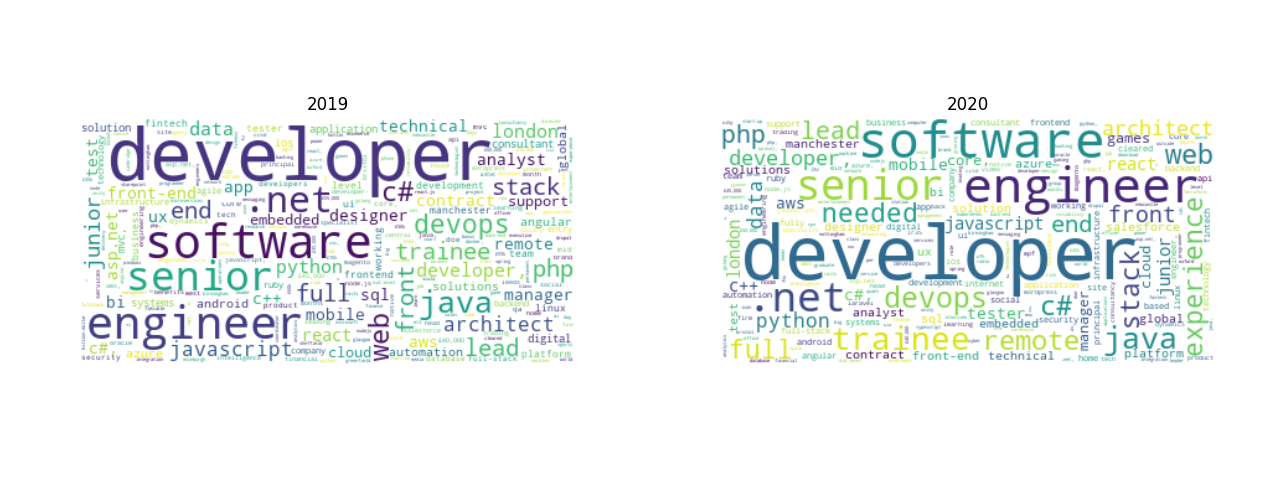}
     \end{subfigure} 
     \begin{subfigure}[H]{\textwidth}
         \centering
         \includegraphics[width=\textwidth]{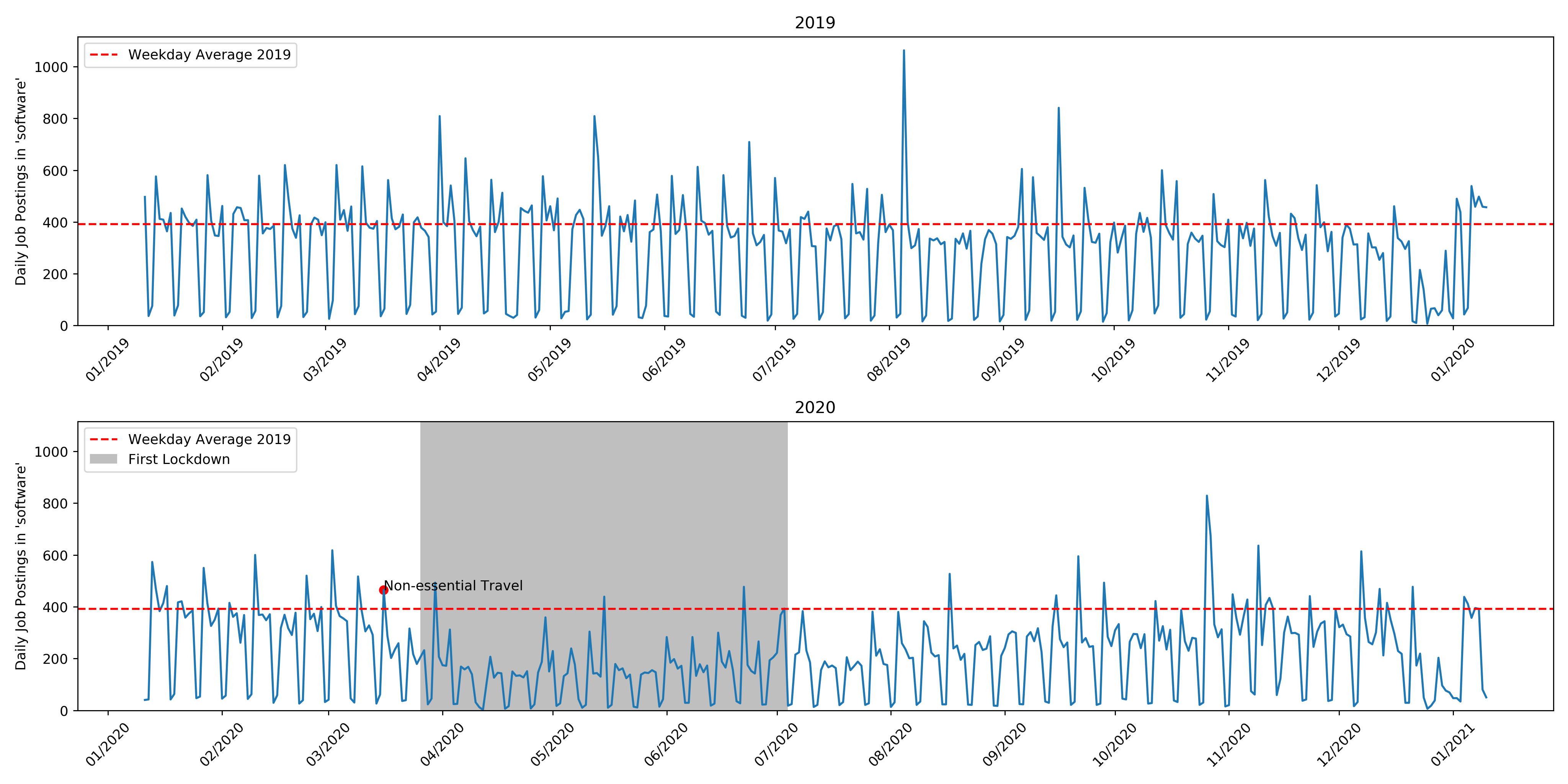}
     \end{subfigure}
        \caption{Jobs matching `software'. The number of these vacancies is down significantly from the pre-crisis average, with the post lockdown weeks suggesting a slow recovery. Many software development jobs can, in principle, be done remotely but the wider economic uncertainty likely affects the number of vacancies companies are willing to advertise. However in the 2020 word cloud we can see the term `remote', an indication that remote working in this industry is becoming widespread. }
        \label{fig:software}
\end{figure}

\begin{figure}[H]
     \centering
     \begin{subfigure}[H]{\textwidth}
         \centering
         \includegraphics[width=\textwidth]{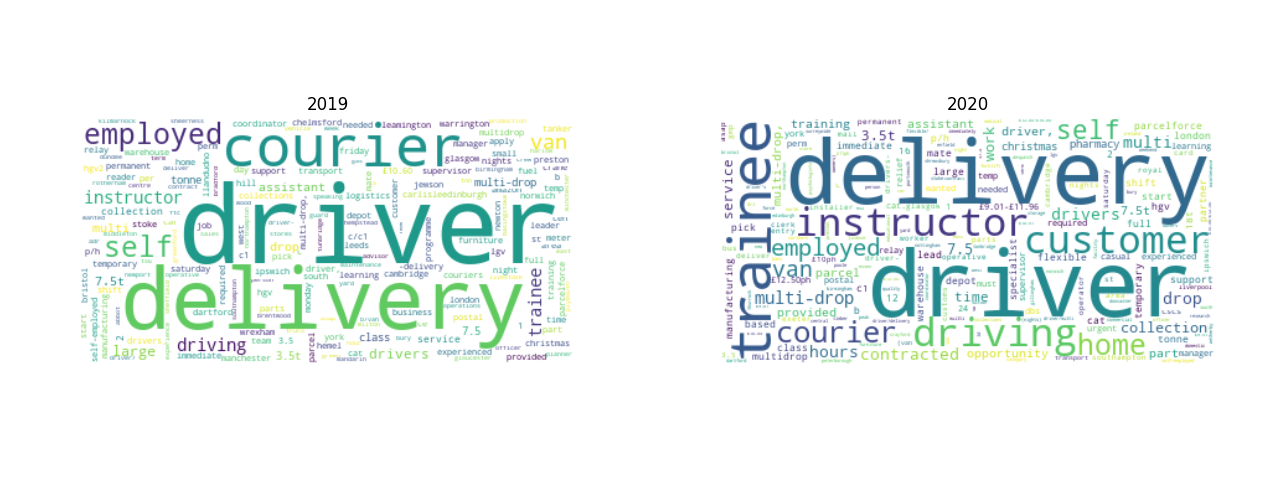}
     \end{subfigure} 
     \begin{subfigure}[H]{\textwidth}
         \centering
         \includegraphics[width=\textwidth]{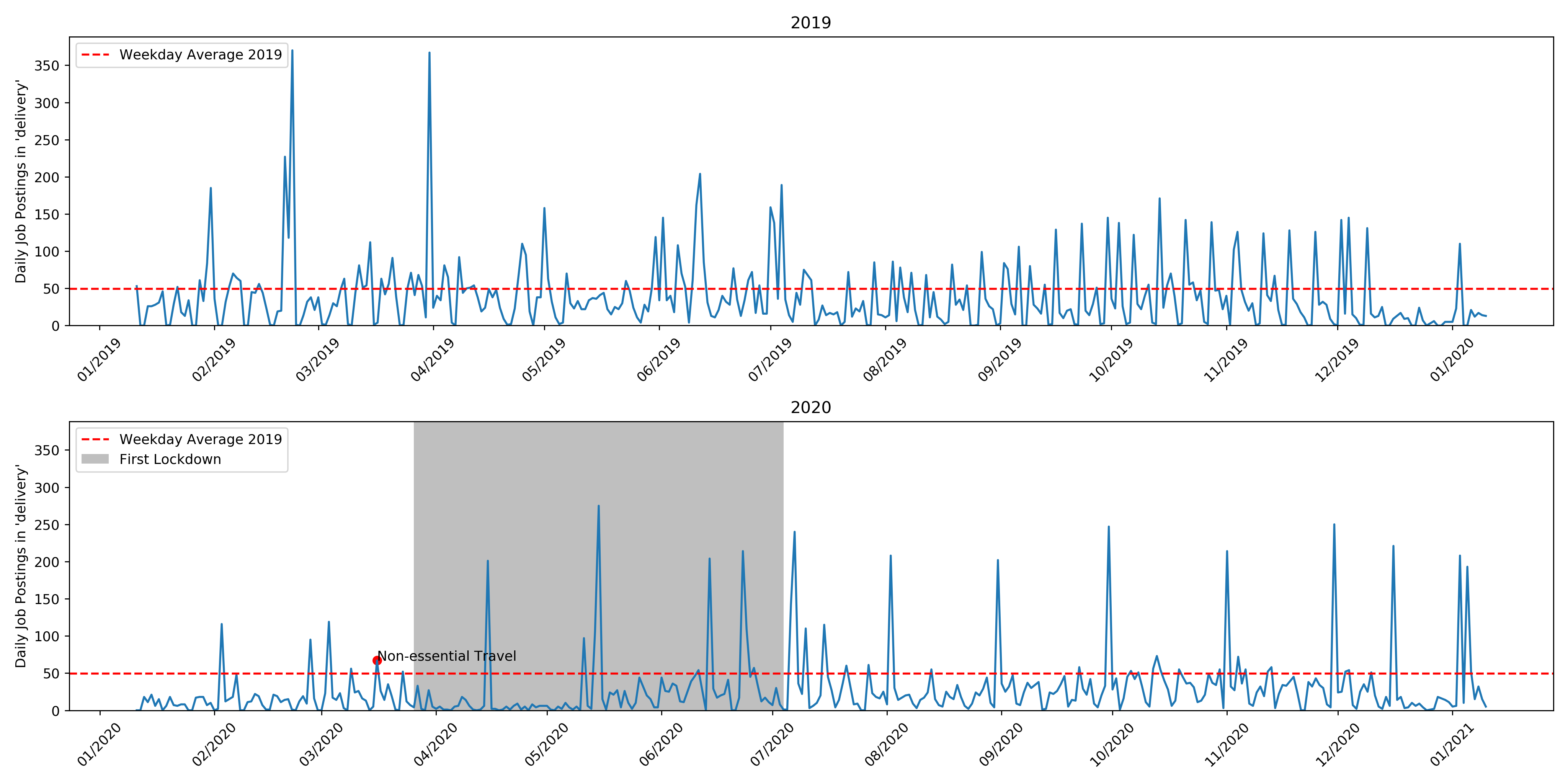}
     \end{subfigure}
        \caption{Jobs matching `delivery'. With most high street stores closed throughout lockdown the number of deliveries could reasonably be expected to increase. However, at the beginning of lockdown the number of adverts for drivers decreased significantly, in line with trends in other sectors. With the easing of restrictions some of this backlog was filled and the number of adverts  is now above pre-crisis levels. This unexpected behaviour shows that companies reacted to lockdown by reducing costs, rather than hiring in anticipation of increased demand. In late 2020 the spikes at the end of every month starting in August are 200 identical adverts for trainee driving instructors that appear to have been misclassified as they contain many terms relating to driving. The total number of these jobs seems to be very low, it is therefore likely that Reed is not a popular platform to advertise this type of job. }
        \label{fig:delivery}
\end{figure}

\section{Jobs by Location}
The regions of the UK were not affected equally by COVID-19. As of January 27th the death rates per 100,000 in the North East and North West were 198.7 and 203.4 respectively, while the rate in the South West was only 89.6 \cite{deaths}. After the first lockdown in some areas with high case rates a local lockdown was imposed. The first of these was the city of Leicester and surrounding areas on July 4th \cite{leicester}, with another significant local lockdown implemented on July 25th in the area of Blackburn with Darwen \cite{blackburn}. Large areas of the North of England were subsequently subjected to more severe restrictions than the rest of the UK \cite{uklockdownnorth}. In this section we will examine how the effect of the crisis on vacancies was distributed across the UK; we will compare regions which were affected to different degrees by COVID-19 and examine if local lockdowns have a compounding effect on depressing the job vacancy data. Below we show time series for different geographic regions, with commentary in the captions.

\begin{figure}[H]
    \centering
    \includegraphics[width=\textwidth]{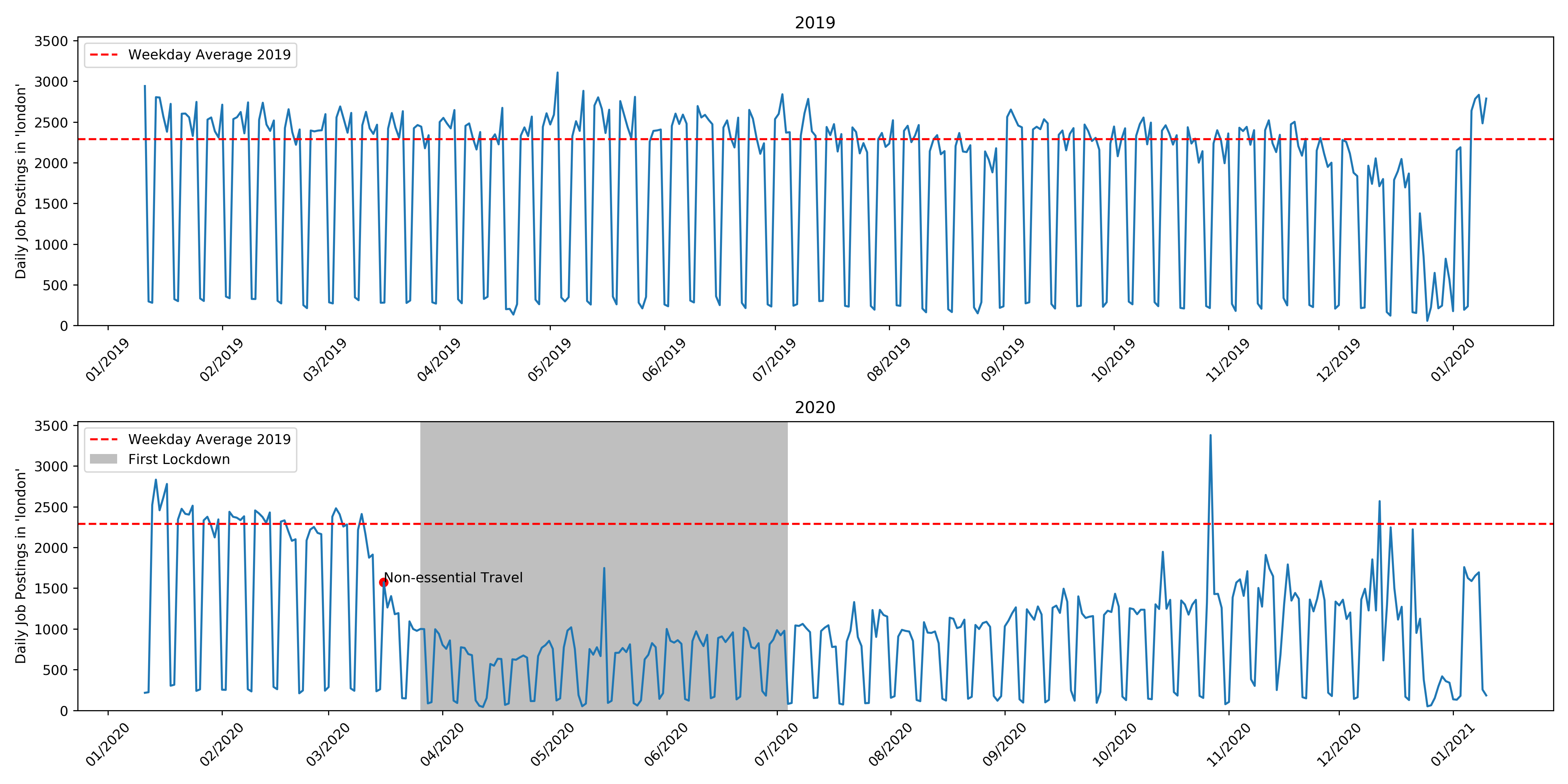}
    \caption{The number of job adverts in the Greater London area. The pattern for London reflects the pattern for the whole UK shown in Figure \ref{fig:total}. }
    \label{fig:london}
\end{figure}

\begin{figure}[H]
    \centering
    \includegraphics[width=\textwidth]{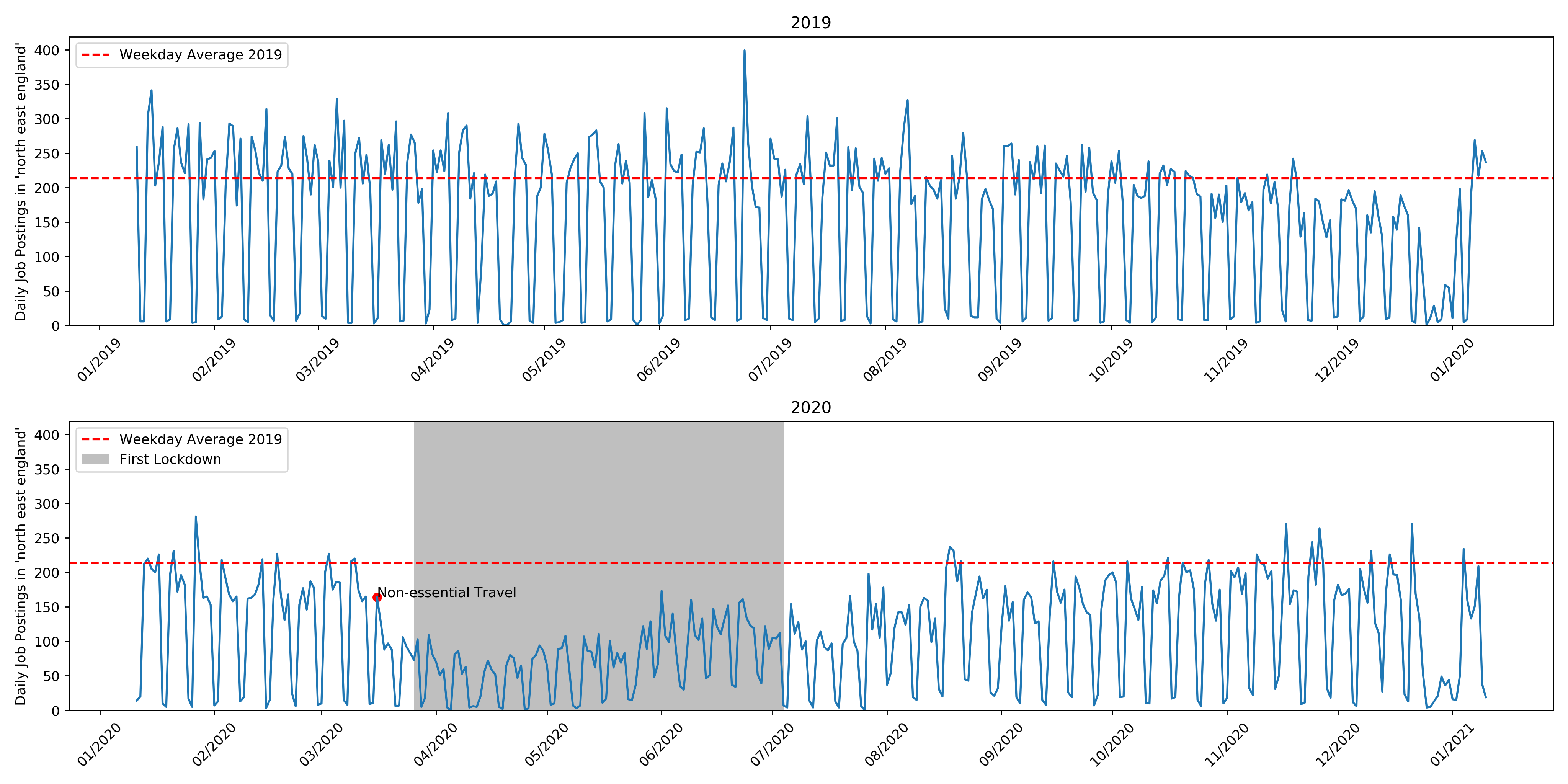}
    \caption{The number of job adverts in the North East. The north east was one of the worst affected regions in the UK. Surprisingly, unlike the pattern for the whole of the UK and Greater London, after the lockdown eased the number of job adverts recovered to near pre-crisis level by mid-August. However, note that the absolute number of jobs is significantly lower, at around 200 postings per day. }
    \label{fig:ne}
\end{figure}

\begin{figure}[H]
    \centering
    \includegraphics[width=\textwidth]{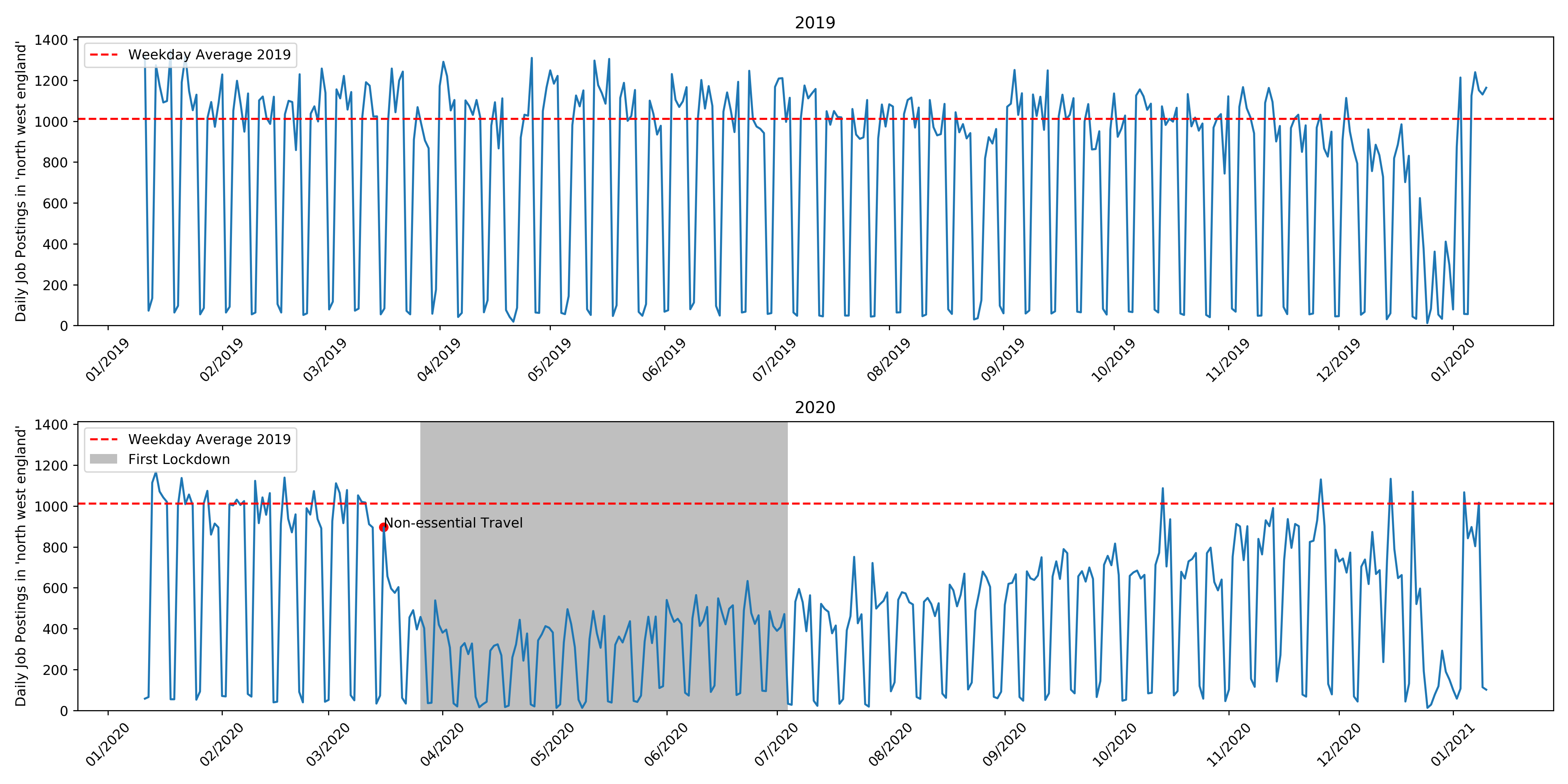}
    \caption{Showing the number of job adverts in the North West. The North West was the worst affected region in the UK and subject to different rules after August 5th \cite{uklockdownnorth}. Despite this, the time series looks broadly similar to the one for Greater London, Figure \ref{fig:london}, and the UK as a whole, Figure \ref{fig:cleantotal}. }
    \label{fig:nw}
\end{figure}

\begin{figure}[H]
    \centering
    \includegraphics[width=\textwidth]{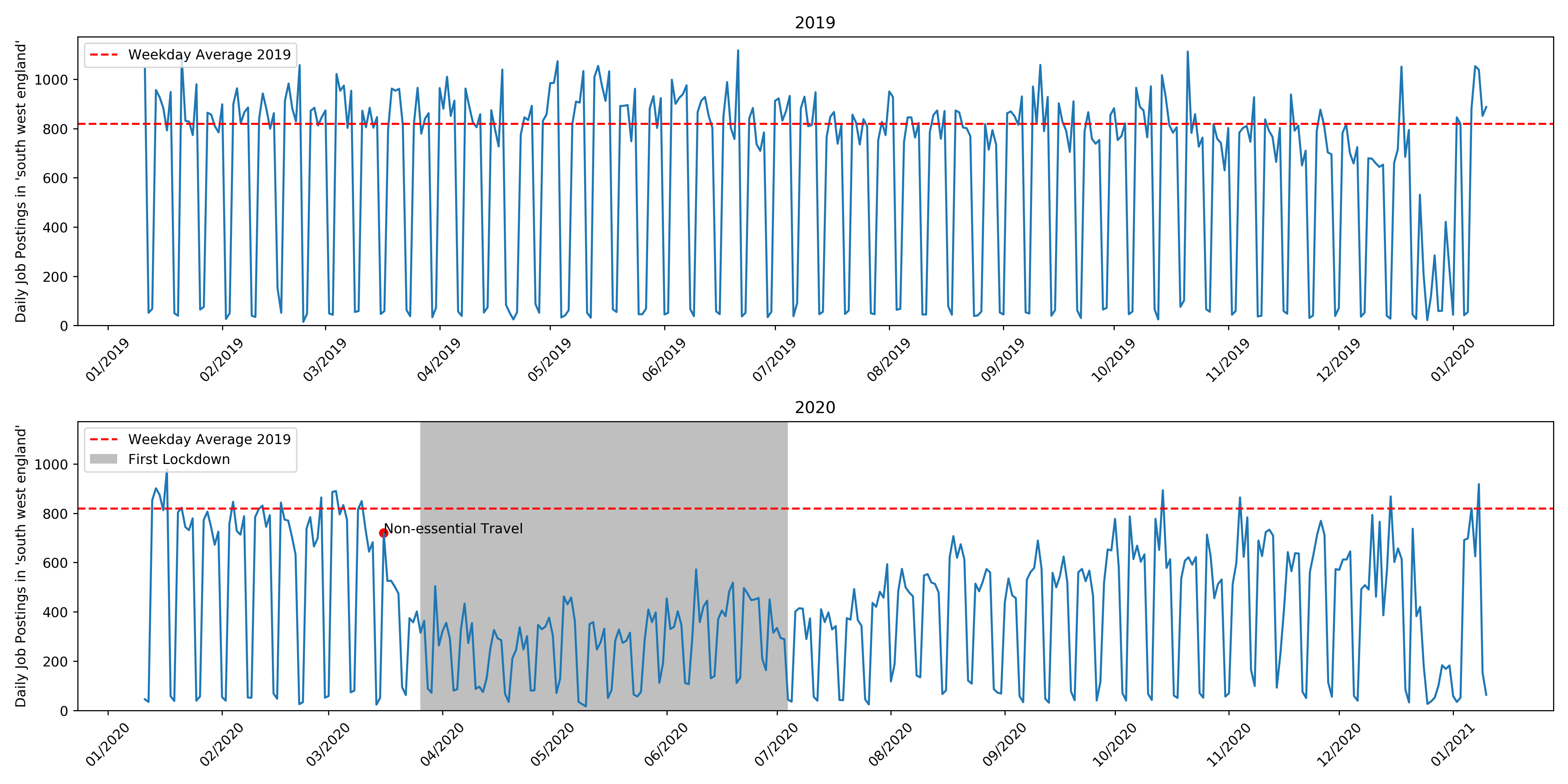}
    \caption{Showing the number of job adverts in the South West. The South West was the region of England least affected by COVID-19.  By mid-September the number of job vacancies recovered slightly more than the UK average, but the drop in job adverts and the slow recovery after lockdown easing reflects the pattern for the whole of the UK shown in Figure \ref{fig:cleantotal}.}
    \label{fig:sw}
\end{figure}

\begin{figure}[H]
    \centering
    \includegraphics[width=\textwidth]{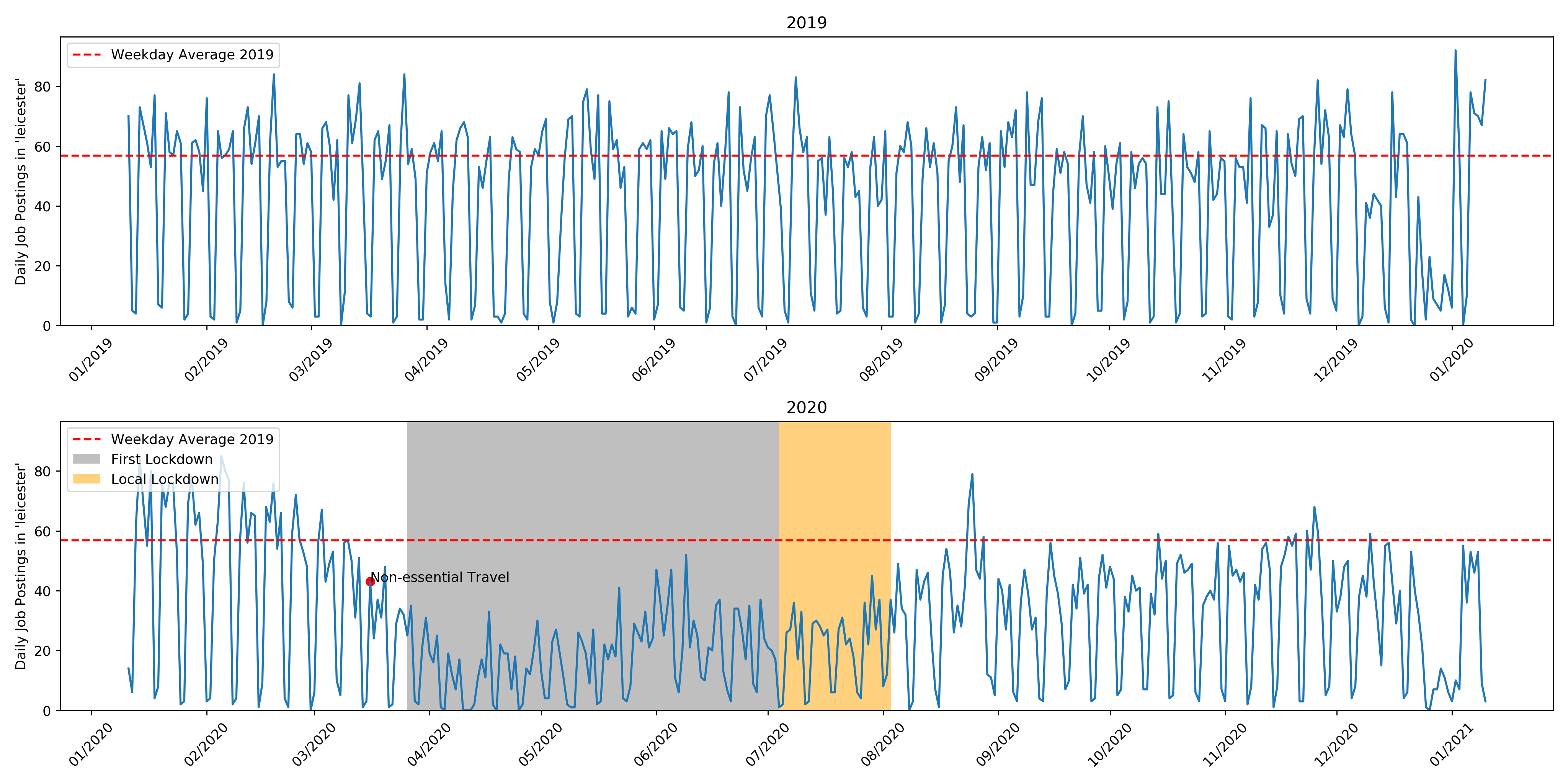}
    \caption{Showing the number of job adverts in the city of Leicester. Leicester saw the first local lockdown, where the rules of the national lockdown period were extended within the city of Leicester and surrounding areas \cite{leicester}. There is no marked change in the number of job adverts in response to this local restriction, with the count staying around its previous (depressed) level. Lifting the local restrictions does correspond to an increase in the number of adverts posted, but this increase is slight and in line with national trends, so determining causality requires further investigation. }
    \label{fig:leicester}
\end{figure}

\begin{figure}[H]
    \centering
    \includegraphics[width=\textwidth]{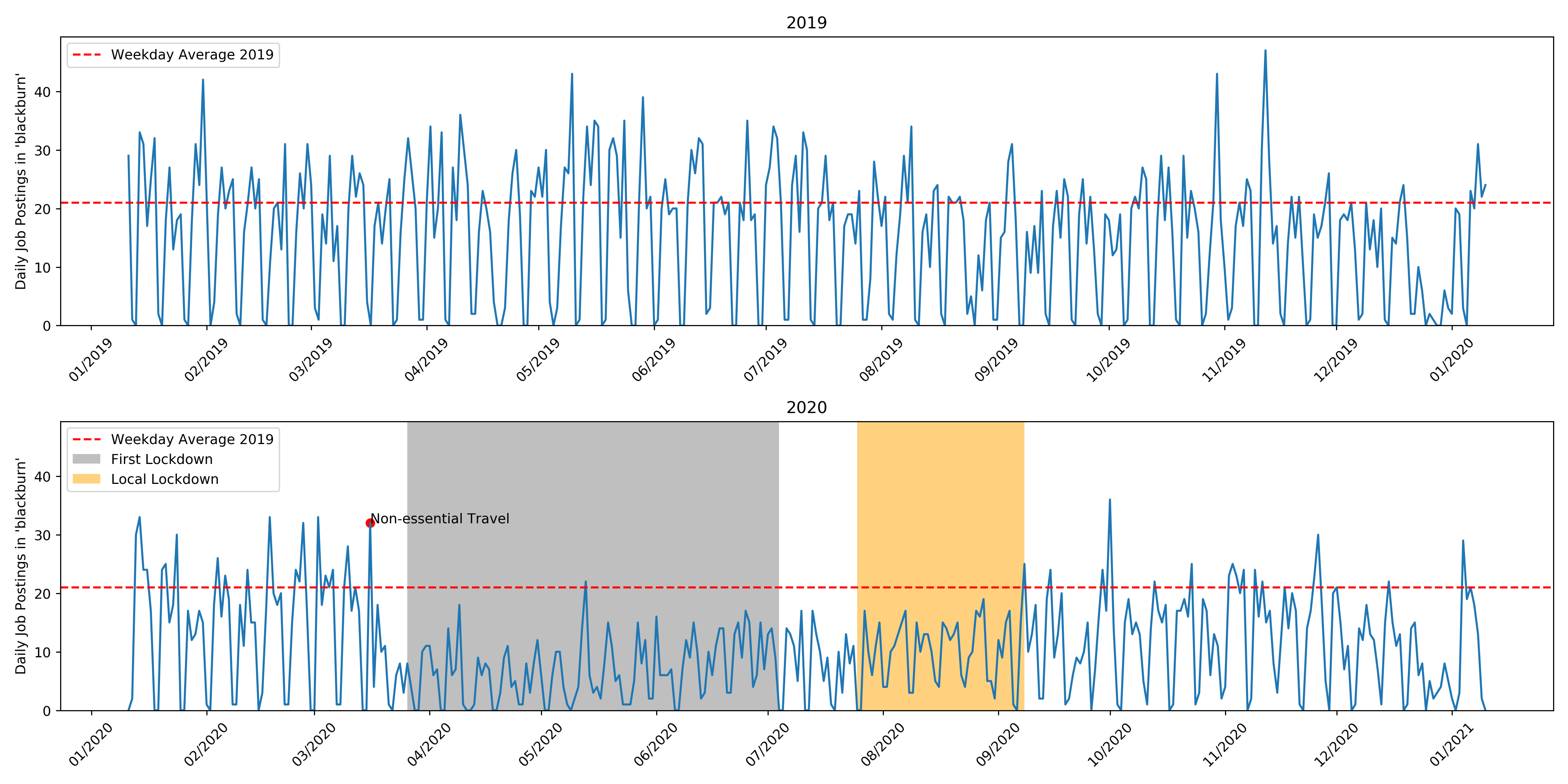}
    \caption{Showing the number of job adverts in area of Blackburn with Darwen. The story is similar to Leicester, show in Figure \ref{fig:leicester}. There is a suggestion that job vacancies increased after the local lockdown. However trends are in line with the rest of the UK and Blackburn is in the north of England and remained subject to different rules \cite{uklockdownnorth} after the local lockdown ended. }
    \label{fig:blackburn}
\end{figure}

\section{Salary and Contract Type}

Other interesting information associated with job adverts includes the salary, contract type (temporary, permanent or contract) and mode of employment (full time or part time). In this section we investigate if there has been any change in the frequency of different contract types, modes of employment or distribution of salaries. We compare the roughly 9 month periods from March 16th 2019 to January 10th 2020 and March 16th 2020 to January 10th 2021, where the second period encompasses the COVID `shock' to the job market.

\begin{figure}
    \centering
    \includegraphics{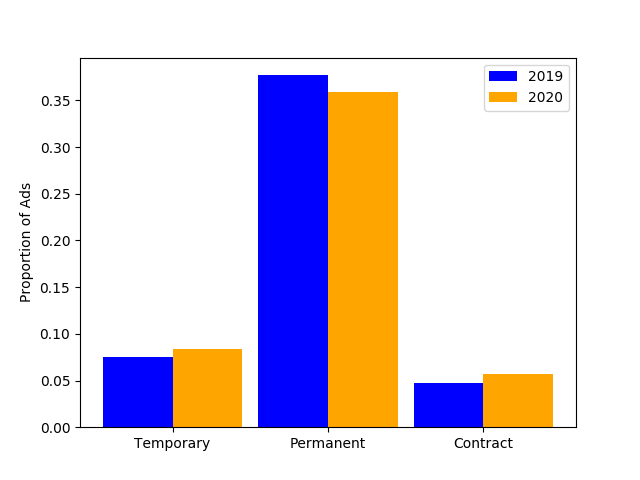}
    \caption{Proportion of job ads corresponding to three types of contract. \textbf{Before Lockdown}: Temporary: 15\%, Permanent: 75\%, Contract 10\%. \textbf{After Lockdown}: Temporary: 17\%, Permanent: 72\%, Contract 11\%. 
    }
    \label{fig:contract}
\end{figure}
\begin{figure}
    \centering
    \includegraphics{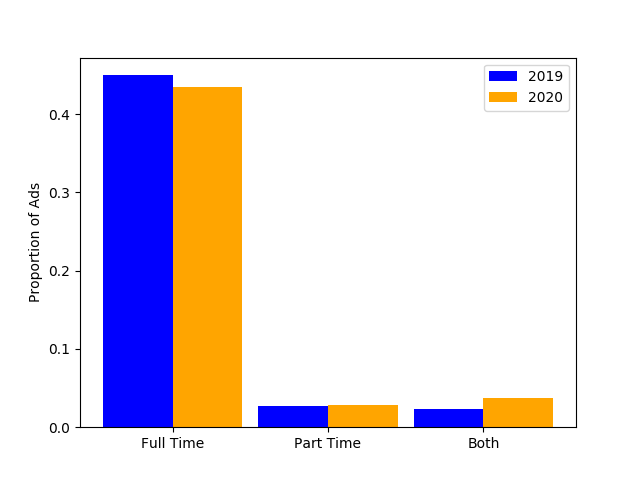}
    \caption{Proportion of job ads corresponding to full time and part time employment. \textbf{Before Lockdown}: Full Time: 90\%, Part Time: 5\%, Both: 5\%. \textbf{After Lockdown}: Full Time: 87\%, Part Time: 6\%, Both: 7\%. }
    \label{fig:mode}
\end{figure}
\begin{figure}
    \centering
    \includegraphics{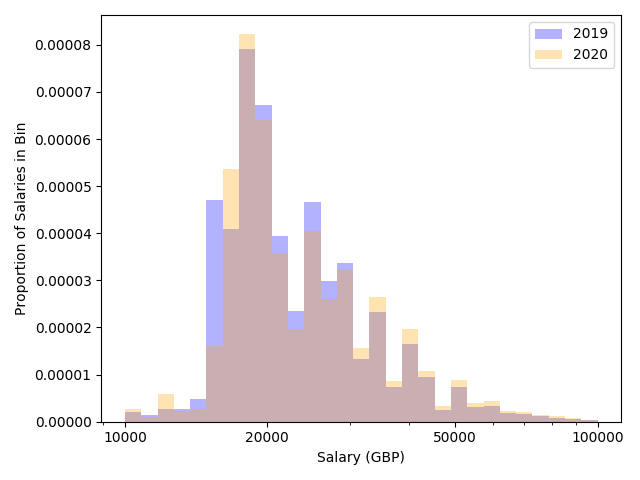}
    \caption{Distribution of annual salaries (obtained from the `yearly minimum salary' field in the JSON returned by the Reed API). \textbf{2019}: Mean Salary: 31285 GBP, Median Salary: 25000 GBP. \textbf{2020}: Mean Salary: 31645 GBP, Median Salary: 25714 GBP.}
    \label{fig:salaries}
\end{figure}

Figure \ref{fig:contract} shows the proportion of jobs advertised as temporary, permanent or contract, Figure \ref{fig:mode} shows the proportion of jobs advertised as full time or part time and Figure \ref{fig:salaries} shows the distribution of advertised salaries. There is a very small increase in mean and median annual salary A t-test comparing the means returns a p-value of 0.167 (no significant difference) while a 2 sample KS-test returns a highly significant (p-value $<10^{-10}$) difference between the two distributions. There is also a slight but significant ($\chi^2$ test p-value $<10^{-10}$) trend towards part time and non-permanent jobs. 

\section{Web Interface}

The topic and geographic filters can be combined to look at e.g. `teaching jobs in the south west' and the salary and contract type analysis can be performed on a sectoral or geographic level. The most effective way to enable stakeholders - e.g. labour market analysts or local authorities - to get this information is likely the creation of an interactive dashboard using the methodology and data described in this work.

\begin{figure}
    \centering
    \includegraphics[width=\textwidth]{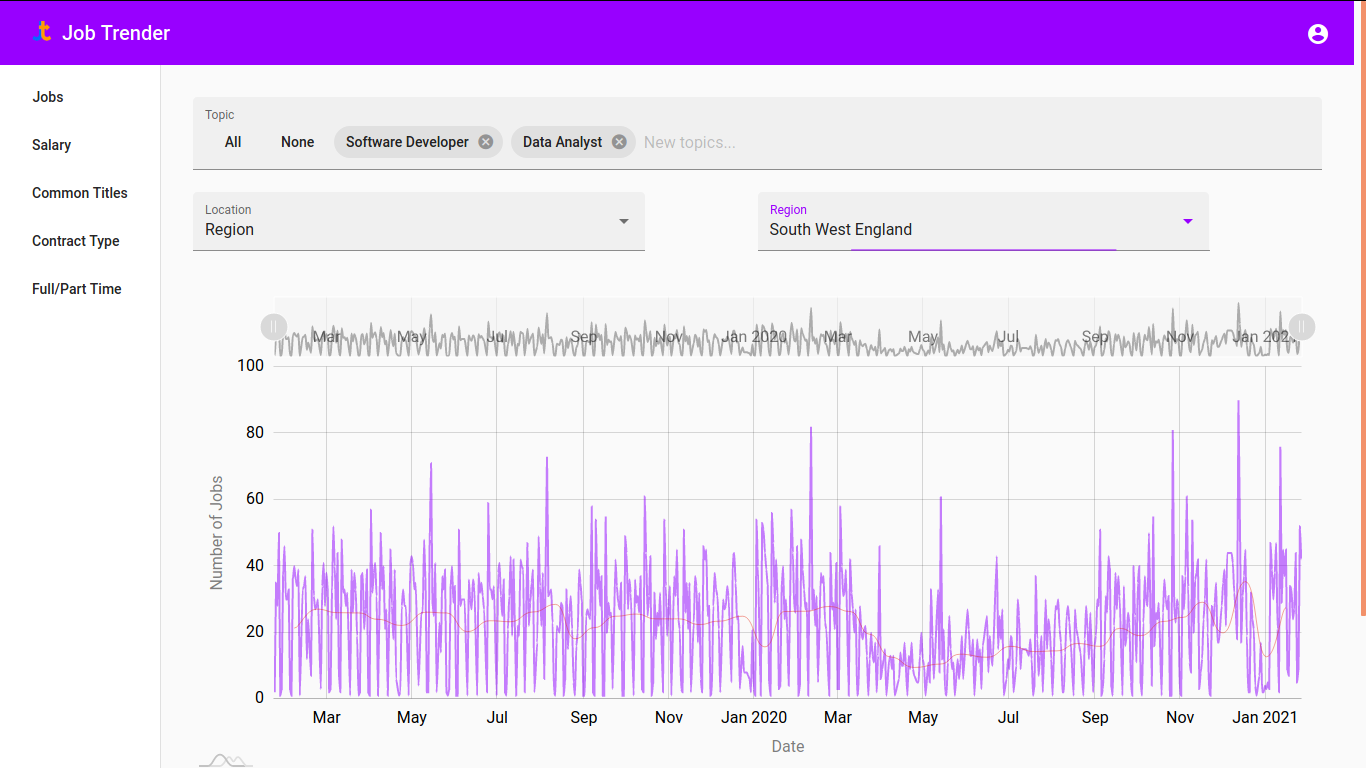}
    \caption{Web view of count of job ads. Shown here are software and data jobs in the south west of England over the entire collection period. }
    \label{fig:web}
\end{figure}
We have created such a dashboard and made it available online at \url{https://jobtrender.com/}. On the `back end' the methods described in this work are used to classify and locate job ads. These are then queried through a simple search interface, an is shown in Figure \ref{fig:web}.

\section{Conclusions}

The data tells us that companies have responded to the COVID-19 crisis by reducing hiring significantly. Certain sectors, like hospitality and graduate recruitment, have been particularly affected  while others, like care work and teaching recruitment, have not been negatively impacted. The conditions of work in a post COVID-19 world, at least in terms of contracts, hours of employment and salary seem to be broadly unchanged but with a very small but statistically significant shift towards higher salary, non-permanent and part time work. One hypothesis is that the lower paying jobs aren't being advertised, increasing the median salary, while the jobs that are being advertised are shifting towards part time and fixed contracts, as companies try to hedge against uncertainty. There may also be seasonal and long term trends. A full understanding of this pattern requires more data.

Regional differences are not as strong as sectoral ones, mirroring the results obtained in the US by \cite{kahn2020labor}. The North East seems to have recovered faster than the national average, but the absolute number of adverts in this area is much lower than in other regions. This could be due to a number of factors: a previously depressed local job market; the local (un)popularity of Reed or even that the types of jobs available in this region not being openly advertised online. Further investigation is required before any strong conclusions can be drawn. It also seems that local lockdowns do not have a compounding effect on the job market but may postpone recovery, though again more investigation is required to fully support this conclusion.

Future work could look at improving our topic detection algorithms, for example using a semi-supervised approach to detect known sectors \cite{ramage2009labeled}. We could also perform topic detection within sectors to study changes in job descriptions in response to new conditions of work e.g. an increase in home working \cite{brynjolfsson2020covid}. 

A major drawback of this work is the use of job adverts from a single source. Though Reed is a large job board there are several others which are as popular in the UK, though they do not allow data collection through an open API. Furthermore there are many specialist recruitment websites e.g. \url{jobs.ac.uk} for academic jobs as well as local ones e.g. \url{devonjobs.gov.uk}. Accumulating job advert information from all of these sources would remove many of the biases (known and unknown) introduced by relying on a single source of data. For example the EU Center for the Development of Vocational Training \cite{european2019online} has attempted to collect online job adverts from across Europe, though they do not reshare their data. What we can do with a single data source is look at \textit{relative} changes in the number of postings, which makes this an effective method for studying the impact of shocks like COVID.

The research of \cite{thurgood2018using} as well as the company Burning Glass shows that using machine learning and geo-inference techniques on job ads can help us to understand labour market trends and skills demand. This research extends those methods and shows they can be used to study short term shocks to the labour market, such as COVID-19 and associated lockdown policies. This is a simple and transparent way to study the effect of economic shocks and major government interventions on labour market activity across employment sectors and geographies. This is complementary data to e.g. survey methodology \cite{chetty2020real} and is accessible to academics, unlike the very useful but opaque methodology of \cite{kahn2020labor} and Burning Glass. 

Though `evidence based policy' is something of a catchphrase \cite{saltelli2017wrong}, it is of course desirable that policy makers consider all the available information before taking action. The effect on the job market should be considered when extending or relaxing lockdown rules and especially in relation to extending or ending furlough and other compensation schemes. We hope that this paper and its methods provide valuable insights for broad or targeted policy interventions in this crisis and the next.

\section*{Supporting information}
\label{sec:SI}
Code available at \url{github.com/rudyarthur/COVIDJobAds}


\nolinenumbers

%
%
%
\bibliographystyle{elsarticle-num-names}
\bibliography{main.bib}


\end{document}